\documentclass{wiley-prop}

\usepackage{algorithm, algorithmicx, algpseudocode}
\usepackage{listings}%
\usepackage{alltt}%
\usepackage{balance}

\articletype{Article Type}%

\received{}
\revised{}
\accepted{}
\doi{10.1002/prop.2019xxxxx}

\newcommand{\be}{\begin{eqnarray}}
\newcommand{\ee}{\end{eqnarray}}

\begin{document}

\title{Braneworld Gravity under gravitational decoupling}

\author[1]{P. Le\'on*}

\author[2]{A. Sotomayor**}


\address[1]{\orgdiv{Departamento de F\'isica}, \orgname{Universidad de Antofagasta}, \orgaddress{\state{Antofagasta}, \country{Chile}}}
\address[2]{\orgdiv{Departamento de Matem\'aticas}, \orgname{Universidad de Antofagasta}, \orgaddress{\state{Antofagasta}, \country{Chile}}}

\corres{ *P. Le\'on. \email{pablo.leon@ua.cl} **A. Sotomayor. \email{adrian.sotomayor@uantof.cl}}

\abstract{We study  the Randall-Sundrum gravity under the gravitational decoupling through the minimal geometric deformation approach (MGD-decoupling). We show a family of new black hole solutions as well as new exact interior solutions for self-gravitating stellar systems and we discuss the corresponding matching conditions.}

\keywords{BraneWorld, Gravitational Decoupling, Internal Solutions, Black Hole Solutions}

\maketitle

\section{Introduction}
\setcounter{equation}{0}
\label{intro}

Einstein's theory of General Relativity, despite its great success, has not been be able to explain in a satisfactory way different fundamental aspects of the gravitational interaction like the existence of dark matter and dark energy and the hierarchy problem. Moreover, this theory breaks down at very high energies, which makes it incompatible with the Standard Model of particles.

Among all the known candidates to describe the gravitational interaction beyond General Relativity, there are theories that include  extra dimensions, which take it inspiration in the Superstring or in the M-theory. These theories are  particularly interesting because they can explain some of the fundamental problems of physics.  In fact, one of these theories, proposed by Randall and Sundrum \cite{Randall:1999vf,Randall:1999ee}, is the Randall-Sundrum Braneworld (RSBW), from which it is possible to explain the scale hierarchy problem.

In the most simple RSBW scenario, our observable universe is modeled as a four dimensional hypersurface, known as the 3-brane, embedded in a five dimensional space, usually called the bulk (there are also consistent realization of the Randall-Sundrum models in higher dimensions, such as in the context of string theory, see for example Ref \cite{Cascales:2003wn}).  The novel idea of the RSBW is that all the gauge interactions, described by the Standard Model, are confined to live in the 3-brane while the gravitational interaction can spread in to the five dimensions of the space. For this reason it is possible to explain the fact that the gravitational scale is very low, compared to the Planck scale, a consequence of the fact that only a part of gravitational interaction is in our four dimensional observable universe, while the another part is spread in a fifth dimension that can be very large. For this reason the study of the modifications of the General Relativity in the Randall-Sundrum models, due to the interaction of the 3-brane with the bulk (see for example \cite{2017bhns.work..201T,Javlon:2019bdt,Rayimbaev:2019nqg}), are very important.

Although a covariant formulation of the RSBW theory \cite{Shiromizu:1999wj} is well known, there are many issues that remain unsolved \cite{Figueras:2011gd,Dai:2010jx,Abdolrahimi:2012qi,Kanti:2013lca,Germani:2001du}. One of the reasons behind these problems is the lack of solutions to the complete five dimensional theory, involving the brane and the bulk interaction. However, an approach  that can shed some light on this problem, consists to find solutions to the effective Einstein's field equations in four dimensions and from it to get some information about the complete geometry in five dimensions. 

Besides, it is well known  that the search of analytical solutions to Einstein's equations in General Relativity is a very difficult task and is even more complicate when we consider in particular interior stellar solutions. This is mostly due to the nonlinearity of the resulting field equations. In the context of the Randall-Sundrum models there are additional nonlinear contributions to the effective energy momentum tensor, in four dimensions, coming from the high energy corrections \cite{Shiromizu:1999wj,Maartens:2010ar}. For this reason the task to find exact physically acceptable solutions for the effective Einstein field equations in four dimensions  seems to be almost impossible.

In the last few years, a new method was found, known as minimal geometric deformation decoupling method (MGD-decoupling), which allows us to solve the Einstein  field equations in a very systematic and simple way. MGD method was proposed~\cite{Ovalle:2007bn,Ovalle:2009xk} in the context of the
brane-world and then was used to obtain new black hole solutions in Refs.~\cite{Casadio:2015gea,Ovalle:2015nfa} (for some works on MGD, see  Refs.~\cite{Casadio:2012pu,Ovalle:2013xla,Ovalle:2013vna,Casadio:2013uma},
and also see Refs. \cite{Ovalle:2014uwa, Casadio:2015jva,Cavalcanti:2016mbe,Casadio:2016aum,daRocha:2017cxu,daRocha:2017lqj,Fernandes-Silva:2017nec,Casadio:2017sze,Fernandes-Silva:2018abr,Ovalle:2016pwp,Fernandes-Silva:2019fez}
for some recent applications). 

Nevertheless, the MGD method, despite being quite powerful, is restricted to consider only isotropic matter distributions, which is a restriction that is not present in the MGD-decoupling approach. Also it is important to mention that in MGD method, the decoupling of the equations of the resulting system, which represents the key to generalize the method to others scenarios beyond the RSBW, was not discussed.  

Indeed, the MGD-decoupling approach is very useful to solve the Einstein's equations for energy-momentum tensors of the form

\begin{equation}
    \bar{T}_{\mu \nu} = T_{\mu \nu}+\theta_{\mu \nu},
\end{equation}
because, instead of solving the complete system of equations for the source $\bar{T}_{\mu \nu}$, we first solve Einstein's equations for the primary source and then we solve a system of equations, similar to Einstein's field equations, for the second source $\theta_{\mu \nu}$. Then, by performing a combination of the two solutions we can obtain the solution for the complete system. Indeed the decoupling of the complete Einstein's equations system, which is highly non trivial, is what makes this method so powerful to analyze different scenarios. 

There are many applications of the MGD decoupling method in different works. One of the approach considered in the literature is to obtain analytical solutions (internal or external) of the Einstein's field equations proposing certain condition in the source $\theta_{\mu \nu}$ (see for example \cite{Ovalle:2017wqi,
Gabbanelli:2018bhs,Ovalle:2018umz,Sharif:2018toc,Morales:2018nmq,Heras:2018cpz,Maurya:2019wsk}). In these cases, the anisotropy generated by the new source can be interpreted in many ways, depending on the main goal of each author, and in general it will be different from the one considered here, which came from the contributions of the bulk into the brane.    
Other solutions obtained using the MGD decoupling method come from the analysis of theories beyond GR like $f(R)$, $f(R,T)$ and $f(G)$ gravity \citep{Sharif:2018tiz,Sharif:2019zan,Maurya:2019hds}, pure Lovelock gravity \cite{Estrada:2019aeh}, AdS/CFT \cite{daRocha:2019pla}, for mention some of them, where the source $\theta_{\mu \nu}$ has a different interpretation with the presented here. For more recent applications of the MGD-decoupling 
method \cite{MGD-decoupling,Ovalle:2019qyi} see for instance Refs.~\cite{Contreras:2018gzd,Contreras:2018vph,Panotopoulos:2018law,Contreras:2019iwm,
Contreras:2019fbk,Ovalle:2018ans,Sharif:2018khl,Morales:2018urp,Sharif:2018pzr,Ovalle:2018vmg,
Gabbanelli:2019txr,Ovalle:2019lbs,Cedeno:2019qkf,Hensh:2019rtb,Contreras:2019mhf,Contreras:2018nfg,Estrada:2018vrl,
PerezGraterol:2018eut,Estrada:2018zbh,Heras:2019ibr,Casadio:2019usg}.

It is the purpose of this work to show how we can use this new method to solve the novelties which appear in the Randal-Sundrum theory in this context. Indeed, we will use this algorithm to find new analytical physically acceptable solutions to the effective Einstein's field equations by extending every known solution of General Relativity  to its braneworld versions. 

We would like to emphasize that while it is true that the braneworld has a well-deserved theoretical importance, it is fair to mention that there is no experimental evidence on it. However, given that it manages to explain the problem of the hierarchy of fundamental forces in a simple and highly non-trivial way, its theoretical importance continues to be recognized until today. Especially since it could serve as a guide to construct a new theory that cannot only explain the problem of the hierarchy of fundamental interactions, but can also be tested experimentally. If this will be possible the MGD-decoupling would be an ideal approach to study the equations of movement  associated with the corrections suffered by General Relativity.

This work is organized as follows: in section 2 we present a brief review of how to decouple Einstein's field equations, using the MGD decoupling method, in the case we have a combination of two gravitational sources. In section 3 we write the four dimensional effective Einstein's equations in the five-dimensional RSBW. In section 4 we explain MGD decoupling in the RSBW domain, giving some black holes solutions; we show also how to find internal solutions for the effective Einstein's field equations in RSBW, taken as a seed the Tolman IV solution in General Relativity; finally, we give a new solution starting with a known solution in the anisotropic domain.
In section 5 we give our conclusions.

\section{Minimal geometric deformation decoupling method}

In this section we present a brief review of the most important results concerning the solution Einstein's equations using the minimal geometric deformation decoupling method for spherically symmetric and static systems.

Let us start by writing the Einstein field equations

\begin{equation}
\label{2.1}
R_{\mu\nu}-\frac{1}{2}\,R\, g_{\mu\nu}=-k^2\bar{T}_{\mu\nu},
\end{equation}
where we will assume that the energy momentum tensor $\bar{T}_{\mu \nu}$ has contributions  of two different gravitational sources, that is

\begin{equation}
\bar{T}_{\mu \nu}=T_{\mu \nu} + \theta_{\mu \nu}.
\label{2.2}
\end{equation}

The line element for our case has the following form in Schwarzschild-like coordinates

\begin{equation}
ds^{2}=e^{\nu (r)}\,dt^{2}-e^{\lambda(r)}\,dr^{2}-r^{2}\left( d\theta^{2}+\sin ^{2}\theta \,d\phi ^{2}\right)
\label{2.3}
\end{equation}
from which we can see in a straightforward way, using (\ref{2.2}), that Einstein's field equations (\ref{2.1}) can be rewritten as

\begin{eqnarray}
\label{2.4}
k^2 \bar{\rho} = k^2 (T^0_0 + \theta_0^0)  &=&\frac{1}{r^2}-e^{-\lambda}\left(\frac{1}{r^2}-\frac{\lambda'}{r}\right)\ ,\\
\label{2.5}
k^2 \bar{p}_r = - k^2 (T^1_1 + \theta_1^1) &=&-\frac 1{r^2}+e^{-\lambda}\left( \frac 1{r^2}+\frac{\nu'}r\right)\ ,\\
\label{2.6}
k^2 \bar{p}_t = -k^2 (T^2_2 + \theta_2^2) &=&\frac{e^{-\lambda}}{4}\Bigg( 2\nu''+\nu'^2- \lambda' \nu' \nonumber \\ & + & \left. 2\frac{\nu'-\lambda'}{r}\right),
\end{eqnarray}
where the prime indicates derivatives respect to variable $r$ and $\bar{\rho}$, $\bar{p}_r$ and $\bar{p}_t$ are defined as the effective energy density, the effective radial pressure and the effective tangential pressure, respectively.

The conservation equation for this system, which can be obtained as a linear combination of the equations (\ref{2.4})-(\ref{2.6}), is given by

\begin{equation}
\nabla_\mu \bar{T}^{\mu \nu} = (\bar{p}_r)'-\frac{\nu'}{2}(\bar{\rho}+\bar{p}_r)-\frac{2}{r}(\bar{p}_t-\bar{p}_r) =0,
\label{2.8}
\end{equation}
which in terms of the gravitational sources $T_{\mu \nu}$ and $\theta_{\mu \nu}$, takes the following form
\begin{eqnarray}
\label{2.9}
\ & & \left({T}_1^{\ 1}\right)'-\frac{\nu'}{2}\left({T}_0^{\ 0}-{T}_1^{\ 1}\right)-
\frac{2}{r}\left({T}_2^{\ 2}-{T}_1^{\ 1}\right)
\nonumber \\ &+& \alpha \left[\left({\theta}_1^{\ 1}\right)'-\frac{\nu'}{2}\left({\theta}_0^{\ 0}-{\theta}_1^{\ 1}\right)-\frac{2}{r}\left({\theta}_2^{\ 2}-{\theta}_1^{\ 1}\right)
\right]=0.
\end{eqnarray}

At this point it is easy to see, from equations (\ref{2.4})-(\ref{2.6}), that the combination of the two sources in the energy-momentum tensor will describe a fluid with local anisotropy on the pressures.

Now in order to solve the system of equations (\ref{2.4})-(\ref{2.6}), we will apply the MGD decoupling method. The first step in this approach is to neglect the contributions of the source $\theta_{\mu \nu}$ and consider a solution of Einstein's equations for the source $T_{\mu \nu}$, whose line element can by written as

\begin{equation}
ds^{2}=e^{\xi (r)}\,dt^{2}-\frac{1}{\mu(r)}\,dr^{2}-r^{2}\left( d\theta^{2}+\sin ^{2}\theta \,d\phi ^{2}\right),
\label{2.10}
\end{equation}
were

\begin{equation}
\label{2.11}
\mu(r)\equiv 1-\frac{k^2}{r}\int_0^r x^2\,T^0_0\, dx
=1-\frac{2\,m(r)}{r},
\end{equation}
is the standard definition of the mass function in General Relativity.

The next step is to include all the contributions  $\theta_{\mu \nu}$ on $T_{\mu \nu}$. This can be done by considering that the effects, induced by the gravitational source $\theta_{\mu \nu}$, are encoded in a deformation of the temporal and radial components of the metric given by

\begin{eqnarray}
\label{2.12}
\nu(r) & = & \xi(r)+ \alpha g(r), \\
\label{3.5}
e^{-\lambda(r)}  & = & \mu(r)+ \alpha f(r) ,
\end{eqnarray}
were $g(r)$ and $f(r)$ are two unknown functions. The minimal geometric deformation is associated with the case were $g=0$. In this particular case we will only be considering deformations in the radial component of the metric, this is

\begin{eqnarray}
\label{2.13}
\nu(r) & = & \xi(r), \\
e^{-\lambda(r)}  & = & \mu(r)+ \alpha f^*(r).
\end{eqnarray}

Now, using (\ref{2.13}), it is easy to show that Einstein's field equations can be separated into two different systems of equations. The first one system corresponds to the Einstein field equations for the source $T_{\mu \nu}$, given by

\begin{eqnarray}
\label{2.14}
k^2 T_0^0 & = & \frac{1}{r^2} -\frac{\mu}{r^2} -\frac{\mu'}{r}\ , \\
\label{2.15}
-k^2 T_1^1 & = & -\frac 1{r^2}+\mu\left( \frac 1{r^2}+\frac{\nu'}r\right)\ , \\
\label{2.16}
-k^2 T_2^2 & = & \frac{\mu}{4}\left(2\nu''+\nu'^2+\frac{2\nu'}{r}\right)+\frac{\mu'}{4}\left(\nu'+\frac{2}{r}\right) \ ,
\end{eqnarray}
with the correspondent conservation equation
\begin{eqnarray}
\label{2.17}
\left({T}_1^{\ 1}\right)'-\frac{\nu'}{2}\left({T}_0^{\ 0}-{T}_1^{\ 1}\right)-
\frac{2}{r}\left({T}_2^{\ 2}-{T}_1^{\ 1}\right) = 0,
\end{eqnarray}
while the second one system is only related to the gravitational source $\theta_{\mu \nu}$ and is written as

\begin{eqnarray}
\label{2.18}
k^2 \theta_0^{\,0}
&\!\!=\!\!&
-\frac{\alpha f^{*}}{r^2}
-\frac{\alpha f^{*'}}{r}
\ ,
\\
\label{2.19}
k^2 \theta_1^{\,1}
&\!\!=\!\!&
-\alpha f^{*}\left(\frac{1}{r^2}+\frac{\nu'}{r}\right)
\ ,
\\
\label{2.20}
k^2 \theta_2^{\,2}
&\!\!=\!\!&
-\alpha \frac{f^{*}}{4}\left(2\nu''+\nu'^2+2\frac{\nu'}{r}\right)
-\alpha \frac{f^{*'}}{4}\left(\nu'+\frac{2}{r}\right)
\ ,
\end{eqnarray}
and the conservation equation associated with the source is
\begin{eqnarray}
\label{2.21}
\left(\theta_1^{\,\,1}\right)'
-\frac{\nu'}{2}\left(\theta_0^{\,\,0}-\theta_1^{\,\,1}\right)
-\frac{2}{r}\left(\theta_2^{\,\,2}-\theta_1^{\,\,1}\right)
=
0.
\end{eqnarray}

We notice that the system of equations (\ref{2.18})-(\ref{2.20}), due to a missing term of $1/r^2$ in the first two equations, is not a Einstein system of equations for $\theta_{\mu \nu}$.  However, it is always possible to redefine the components of $ \theta_{\mu \nu} $ in order to include the factor of $ 1/r^2 $ in the system (\ref{2.18})-(\ref{2.20}) and obtain a Einstein system of equations for this source. Also we can notice, from equations (\ref{2.9}), (\ref{2.19}) and (\ref{2.20}), that the interaction between the sources $T_{\mu \nu}$ and $\theta_{\mu \nu}$ is purely gravitational. Then we can conclude that we have decoupled the Einstein field equations.

The fact of having decoupled the Einstein field equations, for the combination of two sources, represents a huge simplification to the problem of finding solutions to the system of equations (\ref{2.4})-(\ref{2.6}).  This is because instead of solving the equations for the complete system, we can solve first the Einstein's equations for the source $T_{\mu \nu}$, (\ref{2.14})-(\ref{2.17}), and determine $\{T_{\mu \nu},\xi,\mu\}$. Then, we can solve the system of equations (\ref{2.18})- (\ref{2.20}) for the source $\theta_{\mu \nu}$  to find $\{\theta_{\mu \nu},f^*\}$. Finally, the solution for the complete system can be obtained by a simple combination of these two results.

The simple and systematic approach of the MGD decoupling method to solve the Einstein field equations represents a powerful tool in the analysis of more complicated and realistic distributions of matter in the context of General Relativity. Indeed, we can find solutions to the Einstein field equations, for very complicated distributions of matter, in two different ways:

\begin{itemize}
    \item {We can choose known simple solutions of Einstein's equations for the energy-momentum tensor $T_{\mu \nu}$. Then, we can construct more complicated solutions by adding more complex gravitational sources to the energy momentum tensor solving the system (\ref{2.18})-(\ref{2.20}).
    }
    \item {We can start with a very complicated expression for the energy momentum $\bar{T}_{\mu \nu}$. Then in order to find a solution for this case, we can separate the energy-momentum tensor in its more simpler components, that is

    \begin{equation}
      \bar{T}_{\mu \nu} = \sum_i T^i_{\mu \nu}.
    \end{equation}
    Now, we can solve the Einstein's equations for each $ T^i_{\mu \nu}$. Then, by simple combinations of these solutions, we can found the solutions of the Einstein field equations for the more general energy-momentum tensor $\bar{T}_{\mu \nu}$.
    }
\end{itemize}

\section{Field equations in the RSBW}

The main feature of the RSBW models is to consider that our $(3+1)$ observable universe is confined on a 3-brane in a five dimensional space time, usually called the bulk, with $Z_2$ symmetry. This five dimensional theory induces modifications to the Einstein field equations on the brane, which can by written as

\begin{equation}
G_{\mu \nu} = -g_{\mu \nu} \Lambda -k^2 T^T_{\mu \nu},
\label{3.1}
\end{equation}
where $k^2=8\pi G_N$, $\Lambda$ is the cosmological constant on the brane and $T^T_{\mu \nu}$ is a effective energy-momentum tensor given by

\begin{equation}
    T^T_{\mu \nu}= T_{\mu \nu} - \frac{6}{\sigma}S_{\mu \nu} +\frac{1}{8\pi} \mathcal{E}_{\mu \nu} +\frac{4}{\sigma}\mathcal{F}_{\mu \nu},
    \label{3.2}
\end{equation}
which, through the inclusion of the last three terms, take into account all the effects of the bulk onto the 3-brane. Here $\sigma$ is the brane tension.

The term $S_{\mu \nu}$, called the high-energy corrections, in the effective energy-momentum tensor arise from the extrinsic curvature terms in the projected Einstein tensor onto the brane. This is given by

\begin{equation}
    S_{\mu \nu}= \frac{1}{12}TT_{\mu \nu}-\frac{1}{4}T_{\mu \rho} T^\rho_\nu + \frac{1}{24} \left[3T_{\rho \lambda} T^{\rho \lambda}-T^2\right], \label{3.3}
\end{equation}
where $T$ is the trace of $T_{\mu \nu}$. The third  term, $\mathcal{E}_{\mu \nu}$, is known by the name of Kaluza-Klein corrections and represents the projection of the Weyl tensor of the bulk. For the case of spherically  symmetric and static distributions of matter, which is the only case that we will consider in this paper, this term can by written as

\begin{equation}
    k^2 \mathcal{E}_{\mu \nu} = \frac{6}{\sigma}\left[\mathcal{U}\left(u_\mu u_\nu -\frac{1}{3}h_{\mu \nu}\right)+\mathcal{P}_{\mu \nu}\right], \label{3.4}
\end{equation}
with

\begin{eqnarray}
h_{\mu \nu} & = & g_{\mu \nu}-u_\mu u_\nu, \label{3.5} \\
\mathcal{P}_{\mu \nu} & = & \mathcal{P}\left(r_\mu r_\nu -\frac{1}{3}h_{\mu \nu}\right), \label{3.6}
\end{eqnarray}
where $\mathcal{U}$, $\mathcal{P}_{\mu \nu}$, $h_{\mu \nu}$, $u_\mu$ and $r_\mu$ are the bulk Weyl scalar, the anisotropic stress, the projection operator operator, the four velocity of fluid element and a radial unitary vector, respectively.

The last correction to the effective energy-momentum tensor, $\mathcal{F}_{\mu \nu}$, depends on all the stresses in the bulk apart from the cosmological constant. Thus, in general there will be a change of energy momentum between the bulk and the brane. From now on, we will consider only the case in which the cosmological constant is present in the bulk. In this particular case, we recover the standard conservation equation of GR

\begin{equation}
    \nabla^\nu T_{\mu \nu}=0. \label{eqq}
\end{equation}

Now, in order to study the effects of the RSBW on perfect fluids the energy-momentum tensor $T_{\mu \nu}$ we must have the following form

\begin{equation}
    T_{\mu \nu}=(\rho + p_t)u_\mu u_\nu -p_tg_{\mu \nu}+(p_r-p_t)s_{\mu}s_{\nu},
    \label{3.7}
\end{equation}
where $u_{\mu}=\exp{\nu/2}\delta^\mu_0$, $s_{\mu}=\exp{\lambda/2}$ and $\rho$,$p_r$,$p_t$ represents the energy density, the radial pressure and the tangential pressure of the perfect fluid, respectively. In this case the equilibrium equation leads to

\begin{equation}
    p_r'+\frac{\nu'}{2}(\rho +p_r)-\frac{2\Delta}{r}=0, \quad \Delta=p_t-p_r. \label{eqpf}
\end{equation}

Finally, with all these ingredients, we are ready to write the effective Einstein's equation, with $\Lambda=0$, in the four dimensional 3-brane. Then, using (\ref{2.10})  and (\ref{3.2})-(\ref{3.7}), the equation (\ref{3.1}) leads to

\begin{eqnarray}
k^2 \Bigg[\rho  & + & \frac{1}{\sigma}\left(\frac{(\rho^2-\Delta^2)}{2}+\frac{6\mathcal{U}}{k^4}\right)\Bigg]  = \nonumber \\ && \frac{1}{r^2} -e^{-\lambda}\left(\frac{1}{r^2}-\frac{\lambda'}{r}\right) , \label{3.8} \\
k^2 \Bigg[p_r &+& \frac{1}{\sigma}\left(\frac{\rho^2}{2}+ \rho p_t + \frac{p_t^2-p^2_r}{2} +\frac{2\mathcal{U}}{k^4}\right) + \frac{4\mathcal{P}}{k^4\sigma}\Bigg]  = \nonumber \\ &-&  \frac 1{r^2}+ e^{-\lambda}\left( \frac 1{r^2}+\frac{\nu'}r\right),  \label{3.9} \\
k^2 \Bigg[p_t &+& \frac{1}{\sigma}\left(\frac{\rho^2}{2}+ \frac{\rho}{2}(p_r+p_t)+\frac{2\mathcal{U}}{k^4}\right) -\frac{2\mathcal{P}}{k^4\sigma}\Bigg]  = \nonumber \\ && \frac{e^{-\lambda}}{4}\left(2\nu''   +  \nu'^2- \lambda' \nu' +2\frac{\nu'-\lambda'}{r}\right) \label{3.10}.
\end{eqnarray}

Now, from these equations, it is evident that the inclusion of the high energy corrections to the Einstein's equations represents a huge complication respect to General Relativity case. Furthermore, it is easy to see that the equations (\ref{3.8})-(\ref{3.10}) represent an indefinite system because extra information is required, related with the geometry of the bulk, in order to solve the system. In next sections we will show how to solve these problems using the MGD-decoupling method.

\section{Gravitational decoupling in the RSBW}
\label{conc10ch10}

\par
In this section we implement the MGD decoupling method to the effective Einstein's equations obtained for the RS Braneworld scenario, that is, Eqs (\ref{3.8})-(\ref{3.10}). Now, from (\ref{3.2}) (with $\mathcal{F}_{\mu \nu}=0$) it is easy to see that the the source $\theta_{\mu \nu}$ in Eq. (\ref{2.2}) is given by 

\begin{equation}
\theta_{\mu \nu }=-\frac{6}{\sigma}  S_{\mu \nu}+\frac{1}{8\pi}\mathcal{E}_{\mu \nu}. \label{thetach10}
\end{equation}

Then, we can use the expressions (\ref{3.3})-(\ref{3.6}) or, in a more simpler way, by a direct comparison of the field equations (\ref{3.8})-(\ref{3.10}) and (\ref{2.4})-(\ref{2.6}) to find 

\begin{equation}
\theta_0^{\ 0}\,=\,\strut \displaystyle\frac{1}{\sigma }\left( \frac{(\rho^2-\Delta^2)}{2}+\frac{6\mathcal{U}}{k^4}\right) \,,  \label{efecdench10}
\end{equation}%
\begin{equation}
\theta_1^{\ 1}\,=\,-\frac{1}{\sigma }\left( \frac{\rho^2}{2}+ \rho p_t + \frac{p_t^2-p^2_r}{2} +\frac{2\mathcal{U}}{k^4}\right) -\frac{4}{k^{4}}\frac{\mathcal{P}}{%
	\sigma }\,,  \label{efecprerach10}
\end{equation}%
\begin{equation}
\theta_2^{\ 2}\,=\,-\frac{1}{\sigma }\left( \frac{\rho^2}{2}+ \frac{\rho}{2}(p_r+p_t)+\frac{2\mathcal{U}}{k^4}\right) +\frac{2}{k^{4}}\frac{\mathcal{P}}{%
	\sigma }\,.  \label{efecpretanch10}
\end{equation}%

Therefore, the complete system of equations (\ref{3.8})-(\ref{3.10}) can be decoupled in two different systems. The first one, related only with the general relativistic anisotropic fluid  part of the energy-momentum tensor, is given by

\begin{eqnarray}
\label{2.142}
k^2 \rho & = & \frac{1}{r^2} -\frac{\mu}{r^2} -\frac{\mu'}{r}\ , \\
\label{2.152}
k^2 p_r  & = & -\frac 1{r^2}+\mu\left( \frac 1{r^2}+\frac{\nu'}r\right)\ , \\
\label{2.162}
k^2 p_t & = & \frac{\mu}{4}\left(2\nu''+\nu'^2+\frac{2\nu'}{r}\right)+\frac{\mu'}{4}\left(\nu'+\frac{2}{r}\right) \ ,
\end{eqnarray}
with its respective conservation equation

\begin{equation}
p_r'+\frac{\nu'}{2}(\rho +p_r)-\frac{2\Delta}{r}=0.
\end{equation}

And the second one, which has all the information related with the effect of the bulk onto the 3-brane, takes the following form (identifying $\alpha=\frac{1}{\sigma}$)

\begin{eqnarray}
\label{ec1dch10f}
&&k^2\,\left( \frac{(\rho^2-\Delta^2)}{2}+\frac{6\mathcal{U}}{k^4}\right)=
-\frac{f^{*}}{r^2}
-\frac{f^{*'}}{r}\ ,
\\
\label{ec2dch10f}
&&k^2\,\left[\left(\frac{\rho^2}{2}+ \rho p_t + \frac{p_t^2-p^2_r}{2} +\frac{2\mathcal{U}}{k^4}\right) +\frac{4\,{\mathcal{P}}}{k^{4}}\right]
= \nonumber \\ && \quad \quad 
\,f^{*}\left(\frac{1}{r^2}+\frac{\nu'}{r}\right)\ ,
\\
\label{ec3dch10f}
&&k^2\,\left[\left( \frac{\rho ^{2}}{2}+\frac{\rho}{2}(p_r+p_t)-\frac{p_r}{2}\Delta +%
\frac{2}{k^{4}}\,\mathcal{U}\right) -\frac{2\,{\mathcal{P}}}{k^{4}}\right]
=
\nonumber
\\
&& \quad \quad \frac{1}{4}
\left[f^{*}\left(2\,\xi''+\xi'^2+2\frac{\nu'}{r}\right)+f^{*'}\left(\nu'+\frac{2}{r}\right)\right]\ ,
\end{eqnarray}
where the conservation equation for this source can be written as

\begin{eqnarray}
\label{conset2ch10xy}
\mathcal{U}' &+&2\mathcal{P}'+\nu'(2\mathcal{U}+\mathcal{P})+\frac{6\mathcal{P}}{r}-\frac{k^4}{2}(\rho'+p_t')(\rho+p_t) \nonumber \\ &-&\frac{k^4}{2r}(\rho+p_t)\Delta -\frac{k^4 \nu'}{4}(\rho^2+p_rp_t+\rho(p_r+p_t)).
\end{eqnarray}
We note that $\Delta=0$ corresponds to the case of a perfect fluid which can be analyzed in the framework of the first version of MGD method.

Its important to mention that in other cases, the coupling alpha has another interpretations. For example, in Ref. \cite{Ovalle:2018ans}, the coupling alpha is a constant that helps to keep track the effect of the energy-momentum tensor associated with a static scalar field which defines a Klein Gordon equation type through a self-interaction potential.

Now, from the last two systems of equations we can see that every  solution in GR (isotropic or anisotropic in pressures) given by the functions $p_r,p_t,\rho,\mu$ and $\nu$ can be extended to the RS braneworld scenario, using this method, if we solve the system of equations (\ref{ec1dch10f})-(\ref{ec3dch10f})  to find the functions $\mathcal{U},\mathcal{P}$ and $f^*$. 

The most simple case which we can analyze with this approach is the extension of the Schwarzschild vacuum solution of the GR to RSBW scenario. In order to do that we only have to impose the condition $p_r=p_t=\rho=0$ in the system (\ref{ec1dch10f})-(\ref{ec3dch10f}). Then, it is possible to find a simple first differential equation for $f^*$ that can by written as

\begin{equation}
\label{tawh}
\left(\frac{\nu'}{2}+\frac{2}{r}\right)(f^{*})'+\left(\nu'' + \frac{(\nu')^2}{2}+\frac{2\nu'}{r}+\frac{2}{r^2}\right)f^*=0
\end{equation}
which solution is given by
\begin{equation}
f^{*}=\frac{D}{F(r)}  \label{jgfu}
\end{equation}
where $D$ is an integration constant and

\begin{equation}
    F(r)=\exp{\int \frac{\nu'' + \frac{(\nu')^2}{2}+\frac{2\nu'}{r}+\frac{2}{r^2}}{\frac{\nu'}{2}+\frac{2}{r}}dr}. \label{intfac}
\end{equation}
Now, using that 
\begin{equation}
e^{-\mu}=e^\nu =1-\frac{2M}{r}
\end{equation}
for the Schwarzschild vacuum solution, we obtain 
\begin{equation}
f^{*}=\frac{D}{2\left(r-\frac{3M}{2}\right)}\,\left(1-\frac{2\,M}{r}\right)\ ,
\end{equation}
then we can write the minimally deformed radial metric component as
\begin{equation}
e^{-\lambda}=\left(1-\frac{2\,M}{r}\right)\left[1+\frac{D}{2 \sigma \left(r-\frac{3M}{2}\right)}\right]\ , \label{nss}
\end{equation}
and the functions

\begin{eqnarray}
\mathcal{U} & = & -\frac{4 \pi  D M}{3 r^2 (3 M-2 r)^2}, \\
\mathcal{P} & = & -\frac{4 \pi  D (4 M-3 r)}{3 r^2 (3 M-2 r)^2} .
\end{eqnarray}

This result represent the only possible deformation of the Schwarzschild exterior vacuum under the MGD-decoupling method in the BW. Nevertheless, this does not represent a new black hole solution for the BW. Indeed, this solution was first found in Ref.~\cite{Germani:2001du} and later on in Ref.~\cite{Casadio:2015jva}. 

There exist another black hole solutions in the BW context  different from the obtained here \cite{Figueras:2011gd,Dadhich:2000am,Alberghi2011}.
The fact that (\ref{nss}) is the only possible deformation to the Schwarzschild vacuum  shows that the  MGD-decoupling method has strong limitations to obtain new black holes solutions in the RSBW.  Thus, in order to obtain new black solutions we can follow two different approaches. The first one is to use the extended version of the MGD decoupling method found in \cite{Ovalle:2019qyi} and in this way try to obtain more general deformations to the Schwarzschild vacuum. The second one procedure that we could use is based in the standard version of the MGD decoupling method, but instead of using the Schwarzschild vacuum, we can start with a solution in which the vacuum have other contributions coming from different sources from ${\cal U}$ and ${\cal P}$ (see for example Ref.~\cite{Ovalle:2018umz}). In this paper we will only follow the second approach.

\subsection{Black holes by MGD-decoupling in the RSBW}
\par

In order to find new black hole solutions in the RSBW, we suppose that the energy-momentum tensor (\ref{2.2}) is given by 

\begin{equation}
\label{emtch102}
T^{\rm (tot)}_{\mu\nu}
=\frac{6%
}{\sigma }\,S_{\mu \nu }+\frac{1}{8\pi }\,\mathcal{E}_{\mu \nu }
+\,\theta_{\mu\nu}
\ ,
\end{equation}
where the first two terms are  related with BW sector and $\theta_{\mu \nu}$ is another source filling the vacuum. Now if we applied the MDG decoupling method to this system it is easy to see that the two resulting equations systems are given by

\begin{eqnarray}
\frac{6\mathcal{U}}{\sigma k^2} & = &  \frac{1}{r^2} -\mu \left(\frac{1}{r^2}-\frac{\lambda'}{r}\right) , \label{3.82} \\
\frac{2\mathcal{U}}{\sigma k^2} + \frac{4\mathcal{P}}{k^2\sigma} &  = & - \frac 1{r^2}+ \mu \left( \frac 1{r^2}+\frac{\nu'}r\right),  \label{3.92} \\
\frac{2\mathcal{U}}{\sigma k^2} -\frac{2\mathcal{P}}{k^2\sigma} &  = & \frac{\mu}{4}\left(2\nu''   +  \nu'^2- \lambda' \nu' +2\frac{\nu'-\lambda'}{r}\right) \label{3.102},
\end{eqnarray}
which has all the information about the pure BW sector and 

\begin{eqnarray}
\label{2.182}
k^2 \theta_0^{\,0}
&\!\!=\!\!&
-\frac{\alpha f^{*}}{r^2}
-\frac{\alpha f^{*'}}{r}
\ ,
\\
\label{2.192}
k^2 \theta_1^{\,1}
&\!\!=\!\!&
-\alpha f^{*}\left(\frac{1}{r^2}+\frac{\nu'}{r}\right)
\ ,
\\
\label{2.202}
k^2 \theta_2^{\,2}
&\!\!=\!\!&
-\alpha \frac{f^{*}}{4}\left(2\nu''+\nu'^2+2\frac{\nu'}{r}\right)
-\alpha \frac{f^{*'}}{4}\left(\nu'+\frac{2}{r}\right)
\ ,
\end{eqnarray}
which is only related with the source $\theta_{\mu \nu}$. 

Now, in order two solve these systems we use the simplest black hole solution of the RSBW, which in known by the name of tidally charge solution (Ref.\cite{Dadhich:2000am})

\begin{equation}
\label{tidalzzz}
ds^2=\left(1-\frac{2\,M}{r}-\frac{Q}{r^2}\right)\,dt^2-\frac{dr^2}{1-\frac{2\,M}{r}-\frac{Q}{r^2}}-r^2\,d\Omega^2\ .
\end{equation}

Thus, using this solution the system (\ref{3.82})-(\ref{3.102}) is automatically solved and we have only to solve the system (\ref{2.182})-(\ref{2.202}) in order to find $\theta_{\mu \nu}$ and $f^*$ and, in this way, we obtain the deformed version of the tidally charge solution 

\begin{eqnarray}
\label{tidalzzzz}
ds^2 & = & \left(1-\frac{2\,M}{r}-\frac{Q}{r^2}\right)\,dt^2-\frac{dr^2}{1-\frac{2\,M}{r}-\frac{Q}{r^2}+\alpha\,f^{*}} \nonumber \\ & - &r^2\,d\Omega^2\ .
\end{eqnarray}

However, the last system of equations for $\theta_{\mu \nu}$ have more unknown functions that equations and hence we need to give an additional equation. In the following we will consider three different restrictions over the components of the $\theta_{\mu \nu}$.

\subsubsection{The isotropic pressure case $\theta^1_1=\theta^2_2=\theta^3_3$}  

\par
By imposing this condition in the system (\ref{2.182})-(\ref{2.202}) it is possible to get the following first order differential equation for $f^*$,
\begin{equation}
\label{giso}
f^{*'}\left(\nu'+\frac{2}{r}\right)
+f^{*}\left(2\,\nu''+\nu'^2-2\,\frac{\nu'}{r}-\frac{4}{r^2}\right)
=
0
\ .
\end{equation}
Now, using (\ref{tidalzzz}) we can obtain that
\begin{equation}
f^*(r)
=
\left(1-\frac{2\,M}{r}-\frac{Q}{r^2}\right)
\left(\frac{r}{a}\right)^2\,e^{\frac{4\,q}{M\,r}}\left(1-\frac{M}{r}\right)^{2+\frac{4\,q}{M^2}}
\ ,
\label{giso2}
\end{equation}
where $a$ is a constant with dimensions of a length. Thus, using the explicit form of $f^*$ it is possible to write the deformed spatial components of the metric by
\begin{eqnarray}
e^{-\lambda} & = &e^\xi + \alpha\,f^* \nonumber \\ 
&=&
\left(1-\frac{2\,M}{r}-\frac{Q}{r^2}\right) \nonumber \\ & \times &
\left[1
+
{\alpha}
\left(\frac{r}{a}\right)^2\,e^{\frac{4\,q}{M\,r}}\left(1-\frac{M}{r}\right)^{2+\frac{4\,q}{M^2}}
\right]
\ .
\end{eqnarray}
However, this solution is not asymptotically flat. Hence if we want to fulfill this condition, the additional  source cannot be isotropic in the pressures.
\subsubsection{The conformal symmetric case $2\theta^2_2 = -\theta^0_0-\theta^1_1$}
\par

In this case, we can also find a first order differential equation for $f^*$ given by 
\begin{equation}
\label{fconf}
f^{*'}\left(\frac{\nu'}{2}+\frac{2}{r}\right)
+
f^{*}\left(\nu''+\frac{{\nu'}^2}{2}+2\,\frac{\nu'}{r}+\frac{2}{r^2}\right)
=
0
\ ,
\end{equation}
and using (\ref{tidalzzz}) we can get that $f^*$ and $e^{-\lambda}$ are also given by  

\begin{equation}
\label{gconf}
f^{*}(r)
=
\left(1-\frac{2\,M}{r}-\frac{Q}{r^2}\right)\frac{b e^{\frac{3MArcTan\left[\frac{3M-4r}{\sqrt{-9M^2-8q}}\right]}{\sqrt{-9M^2-8q}}}}{\sqrt{r\,(2\,r-3\,M)-q}}
\ ,
\end{equation}

\begin{equation}
\label{confsol}
e^{-\lambda}
=
\left(1-\frac{2\,M}{r}-\frac{q}{r^2}\right)
\left(1+\frac{B e^{\frac{3MArcTan\left[\frac{3M-4r}{\sqrt{-9M^2-8Q}}\right]}{\sqrt{-9M^2-8Q}}}}{\sqrt{r\,(2\,r-3\,M)-Q}}\right)
\ ,
\end{equation}
where $B=\sigma^{-1}b$ and $b$ is a constant with units of a length. This is a black hole that exhibits two horizons in $r=M\pm\sqrt{M^2+Q}$, which are the same of the tidal charged solution.
\subsubsection{The null tangential pressure case $\theta^2_2=0$}
\par
In this case it is easy to find that 
\begin{equation}
\label{ec3dch10xx}
f^{*'}\left(\nu'+\frac{2}{r}\right)+f^{*}\left(2\,\nu''+\nu'^2+2\frac{\nu'}{r}\right)=0\ ,
\end{equation}
and introducing (\ref{tidalzzz}) as before, we obtain that $f^*$ and $e^{-\lambda}$ are given by 
\begin{equation}
f^{*}=d\,\left(1-\frac{2\,M}{r}-\frac{Q}{r^2}\right)\left(1-\frac{M}{r}\right)^{\frac{2\,Q}{M^2}}\,e^{\frac{2\,Q}{M\,r}}\ ,
\end{equation}
\begin{equation}
\label{nultang}
e^{-\lambda}
=
\left(1-\frac{2\,M}{r}-\frac{Q}{r^2}\right)
\left[1+d\,\left(1-\frac{M}{r}\right)^{\frac{2\,Q}{M^2}}\,e^{\frac{2\,Q}{M\,r}}\right]
\ ,
\end{equation}
where $d$ is an integration constant. 
This case, as the first one (isotropic pressures), is not asymptotically flat. Thus, we can conclude that if we want to preserve the asymptotically flat property, the source $\theta_{\mu \nu}$ cannot contain a null tangential pressure. 

\subsection{Interior solutions by MGD-decoupling in the RSBW}
\par
Now, let us find interior solutions for a self-gravitating system. The deformed interior metric under MGD-decoupling reads
\begin{equation}
\label{tidalzzz3}
ds^2=\left(1-\frac{2\,m(r)}{r}\right)\,dt^2-\frac{dr^2}{1-\frac{2\,m(r)}{r}+\alpha\,f^{*}}-r^2\,d\Omega^2\ .
\end{equation}
Let us remind that after the decoupling, we end with three independent equations, namely, the system~\eqref{ec1dch10f}-\eqref{ec3dch10f}, to find three unknown functions $\{{\cal U},\,{\cal P},\,f^{*}\}$. By combining Eqs.~\eqref{ec1dch10f}-\eqref{ec3dch10f} we find the first order differential equation for the function $f^{*}$, given by

\begin{eqnarray}
\left(\frac{\xi'}{2}+\frac{2}{r}\right)(f^*)' & + & \left(\xi''+\frac{(\xi')^2}{2}+\frac{2\xi'}{r}+\frac{2}{r^2}\right) f^* \nonumber \\ & = & k^2(\rho^2+\rho(2p_t+p_r)+\Delta^2),
\end{eqnarray}
whose formal solution is

\begin{equation}
\label{tawh2}
f^*(r)=\frac{J}{F(r)}+ \frac{2k^2}{F(r)}\int \frac{F(r)r}{r\xi'+4}(\rho^2+\rho(2p_t+p_r)+\Delta^2)dr,
\end{equation}
with
\begin{equation}
F(r)=\exp{\left(\int\left(\xi''+\frac{(\xi')^2}{2}+\frac{2\xi'}{r}+\frac{2}{r^2}\right)\Big{/}\left(\frac{\xi'}{2}+\frac{2}{r}\right)dr\right)} ,\label{fit}
\end{equation}
where we see that in the vacuum $\rho=p_t=p_r=0$ the expression~\eqref{tawh2} yields the one in~\eqref{tawh}.

Then,  what we have shown here is that given a known solution of Einstein's equations in General Relativity it is always possible to extend it to the Brane World scenario using the MGD-decoupling method. It is important to recall that we are assuming that the source $T_{\mu \nu}$  in (\ref{2.2}) corresponds to a perfect fluid. The case where the $T_{\mu \nu}$ represents a fluid with local anisotropy in pressure can be obtained directly following the same steps that we presented here.

In order to avoid singularities at the surface of our distribution we must impose the well known matching conditions. The exterior geometry in the Brane-Wolrd context is characterized by a Weyl fluid with

\begin{equation}
\label{weylfl}
    \rho=p_r=p_t=0, \quad \mathcal{U}=\mathcal{U}^+, \mathcal{P}=\mathcal{P}^+,
\end{equation}
whose metric can be written in a generic way by

\begin{equation}
    ds^2 = e^{\nu^+} dt^2 -e^{\lambda^+}dr^2 -r^2 d\Omega^2.
\end{equation}

From (\ref{weylfl}) it is easy to see that the effective pressures and the effective density in the outer region will be in general different from zero due to the contributions coming from the interaction of our universe with the bulk. On the other hand, from the Eqs. \eqref{ec1dch10f}-\eqref{ec3dch10f} becomes evident that the system of equations for the exterior region has more unknown functions than equations. So, in order to close the system, it is necessary to impose further conditions under $\mathcal{U}^+$ and $\mathcal{P}^+$. Then, unlike the General Relativistic case, in the BW we can have many possible static and spherically symmetric vacuum solutions, in which the Schwarzschild's vacuum is only a particular case ($\mathcal{U}^+=\mathcal{P}^+=0$). It can be shown that, in the most general case, the first and second fundamental form lead to

\begin{eqnarray}
  \left. e^{\nu^{-}} \right|_{r=R} & = & \left. e^{\nu^{+}}\right|_{r=R} , \label{fff1} \\
  \left.  \left(1-\frac{2m(r)}{r}+ \frac{1}{\sigma}f^*(r)\right)\right|_{r=R}  & = &  \left. e^{-\lambda^+}\right|_{r=R}  , \label{fff2} \\
  \left. \left(p_r(r) + \frac{f^*}{8\pi \sigma}\left[\frac{\nu'}{r}+\frac{1}{r^2}\right]\right) \right|_{r=R} & = & \left. \left(\frac{2}{k^4}\frac{\mathcal{U}^+}{\sigma}+\frac{4}{k^4}\frac{\mathcal{P}^+}{\sigma} \right) \right|_{r=R}. \label{sff}
\end{eqnarray}

In the General Relativistic scenario, the second fundamental form (\ref{sff}) leads to the condition

\begin{equation}
    p_r(R)=0, \label{sffgr}
\end{equation}
however in the BW scenario, even when the physical pressure equal to zero at the surface of the distribution, the effective radial pressure will be different from zero at $r=R$. Now, in this paper we  consider the case when the condition (\ref{sffgr}) is satisfied and also the case when only (\ref{sff}) is fulfilled but not (\ref{sffgr}).
\par

On the other hand, we also will check the physical acceptability of the obtained by requiring the following condition \cite{Delgaty} for the obtained solutions  
 
\begin{itemize}
\item $\bar{p}_r$, $\bar{p}_t$ and $\bar{\rho}$ are positive and finite inside the distribution.
\item $\frac{d\bar{p}_r}{dr}$, $\frac{d\bar{p}_t}{dr}$ and $\frac{d\bar{\rho}}{dr}$ are monotonically decreasing.
\item Dominant energy condition: $\frac{\bar{p}_r}{\bar{\rho}}\leq 1$ \hspace{0.1cm}, \hspace{0.1cm} $\frac{\bar{p}_t}{\bar{\rho}}$ $\leq 1$.
\item Causality condition: $0<\frac{d\bar{p}_r}{d\bar{\rho}}<1$\hspace{0.1cm}, \hspace{0.1cm}$0<\frac{d\bar{p}_t}{d\bar{\rho}}<1$.
\end{itemize}

Now, in order to show how the MGD-decoupling method works let us to extend two known solutions of the Einstein's field equation in GR, one with isotropic pressures and the other with local anisotropic pressures.

\subsubsection{The isotropic case}
For this case we will choose the Tolman IV perfect fluid solution given by

\begin{eqnarray}
e^{\xi} & = & B^2\left(1+\frac{r^2}{A^2}\right), \label{tolman00} \\
e^{-\lambda} & = & \frac{\left(1-\frac{r^2}{C^2}\right)\left(1+\frac{r^2}{A^2}\right)}{\left(1+\frac{2r^2}{A^2}\right)}, \\
\rho(r) & = & \frac{3A^4+A^2(3C^2+7r^2)+2r^2(C^2+3r^2)}{8\pi C^2(A^2+2r^2)^2}, \label{tolmandensity} \\
p(r) & = & \frac{C^2-A^2-3r^2}{8\pi C^2(A^2+2r^2)}, \label{tolmanpressure}
\end{eqnarray}
where $A$, $B$ and $C$ are constants that must be determined by the matching conditions. This perfect fluid configuration $\{p,\,\rho,\,\mu,\,\xi\}$ is a solution of the system~\eqref{2.14}-\eqref{2.16}, and we will have a specific braneworld solution under the MGD-decoupling.  Then, plugging the expressions in Eqs.~\eqref{tolman00}-\eqref{tolmanpressure} in Eq.~\eqref{tawh2}, we found that


\begin{eqnarray}
f^*(r) & = & -\frac{\left(A^2+r^2\right)}{384 \pi  C^4 r \left(2 A^2+3 r^2\right)^{3/2}}\Bigg(r \sqrt{2 A^2+3 r^2} \nonumber \\ &\times & \Bigg( -\frac{9 \left(A^4-4 C^4\right)}{A^2+2 r^2} + \frac{4 A^4 \left(A^2+2 C^2\right)^2}{\left(A^2+2 r^2\right)^3} \nonumber \\ & + & \frac{17 \left(A^3+2 A C^2\right)^2}{\left(A^2+2 r^2\right)^2}- 36 C^2+54 r^2\Bigg) \nonumber \\ &-&  24 \left(A^2+2 C^2\right)^2 \tan ^{-1}\left(\frac{r}{\sqrt{2 A^2+3 r^2}}\right) \nonumber \\  &-& 48 \sqrt{3} C^4 \log \left(\sqrt{6 A^2+9 r^2}+3 r\right) \Bigg) \nonumber \\ & + &  \frac{D(A^2+r^2)}{r \left(2 A^2+3 r^2\right)^{3/2}}.
\end{eqnarray}

from which follows that $f^*$ will be finite at the center of the distribution only if $J=0$ and $A^2=1/6$. This last condition can be seen easily if we expand $f^*(r)$ in a power series around $r=0$, which yields

\begin{eqnarray}
f^*(r) & = & \frac{\sqrt{\frac{3}{2}} \log \left(\sqrt{6} \sqrt{A^2}\right)}{16 \pi  \sqrt{A^2} r} \nonumber \\ &-& \frac{5 r \left(\sqrt{\frac{3}{2}} \sqrt{A^2} \log \left(\sqrt{6} \sqrt{A^2}\right)\right)}{64 \left(\pi  A^4\right)} \nonumber \\ & + & \frac{3 r^2 \left(A^2+C^2\right)}{8 \pi  A^4 C^2} + O(r^3),
\end{eqnarray}
where it is clear that $A^2=1/6$ in order to get an expression for the deformation function finite at the origin. Then $f^*(r)$ can be written as

\begin{eqnarray}
   f^*(r) &=& -\frac{\left(6 r^2+1\right)}{384 \sqrt{3} \pi  C^4 r \left(9 r^2+1\right)^{3/2} \left(12 r^2+1\right)^3} \nonumber \\ &\times & \Bigg(\sqrt{3} \left(r \sqrt{9 r^2+1} \left(72 C^4 \left(216 r^4+70 r^2+5\right) \right. \right.  \nonumber \\ &-& 24 C^2 \left(1296 r^6+324 r^4+10 r^2-1\right) \nonumber \\ &+& \left. 46656 r^8+11664 r^6+864 r^4+26 r^2+1\right) \nonumber \\ &-& \left. 72 C^4 \left(12 r^2+1\right)^3 \log \left(\sqrt{9 r^2+1}+3 r\right)\right)\nonumber \\ &-& \left(12 C^2+1\right)^2 \left(12 r^2+1\right)^3 \tan^{-1}\left(\frac{r}{\sqrt{3 r^2+\frac{1}{3}}}\right)\Bigg). \label{defor}
\end{eqnarray}

Now, using (\ref{defor}) we can found that the functions $\mathcal{U}$ and $\mathcal{P}$ can by written as

\begin{eqnarray}
\mathcal{U} & = &-\frac{4 \pi f^*(r) \left(18 r^2+5\right)}{54 r^4+15 r^2+1} \nonumber \\ &-& \frac{3 \left(6 C^2 \left(4 r^2+1\right)+72 r^4+14 r^2+1\right) G_1(r)}{4 \left(9 r^2+1\right) \left(12 C r^2+C\right)^4}    , \\
\mathcal{P} & = & \frac{2 \pi f^*(r) \left(144 r^4+22 r^2+1\right)}{54 r^6+15 r^4+r^2} \nonumber \\ & + &  \frac{3 r H_1(r)\left(6 C^2 \left(4 r^2+1\right)+72 r^4+14 r^2+1\right) }{2 C^4 \left(9 r^2+1\right) \left(12 r^2+1\right)^3}.
\end{eqnarray}
where
\begin{eqnarray}
    G_1(r) & = & 6 C^2 \left(132 r^4+41 r^2+3\right)-216 r^6 \nonumber \\ &-& 42 r^4+7 r^2+1, \\
    H_1(r) & = & \left(4 \left(9 r^4+r^2\right)-3 C^2 \left(8 r^2+1\right)\right).
\end{eqnarray}

\textbf{The case with $p(R)=0$} \\
Assuming that Eq. (\ref{sffgr}) is satisfied, then  using (\ref{sffgr}) it can be shown that
 \begin{equation}
     C^2 = A^2+3R^2.
 \end{equation}

 Then we can compute the value of the radial pressure at $r=R$, which yields

\begin{equation}
    P(R)= \frac{3(I_1(R) +I_2(R) +I_3(R))}{256 \pi ^2 \sigma R^3 \left(9 R^2+1\right)^{3/2} \left(216 R^4+30 R^2+1\right)},
\end{equation}
where

\begin{eqnarray}
I_1(R) & \equiv & 6 -3 R \sqrt{9 R^2+1} \left(216 R^4+66 R^2+5\right), \\
I_2(R) & \equiv & 3 \sqrt{3} \left(12 R^2+1\right)^3 \tan ^{-1}\left(\frac{R}{\sqrt{3 R^2+\frac{1}{3}}}\right), \\
I_3(R & \equiv & 2 \left(12 R^2+1\right) \left(18 R^2+1\right)^2 \nonumber \\ &\times & \log \left(\sqrt{9 R^2+1}+3 R\right).
\end{eqnarray}
So, in general the radial pressure will be different from zero in the surface of the distribution. Therefore we will not be able to match our internal solution to the Schwarzschild vacuum ($\mathcal{U}^+=\mathcal{P}^+=0$) but it is possible to use the exterior solution found at the beginning of this section, which is a deformation of the Schwarzschild metric. In this case the matching conditions (\ref{fff1})-(\ref{sff}) yield

\begin{eqnarray}
&&B^2 \left(6 R^2+1\right)  =  \left(1-\frac{2M}{R}\right), \label{fffe} \\
&&\frac{6 R^2+1}{18 R^2+1} +\frac{1}{\sigma}f^{*}(R)  =  \left(1-\frac{2M}{R}\right)\nonumber \\ & &  \hspace{3.2cm} \times \left[1+\frac{D}{2\sigma \left(R-\frac{3}{2}M\right)}\right], \label{fffe2} \\
&&\frac{f^*(R)}{8\pi \sigma}\left(\frac{12}{6 R^2+1}+\frac{1}{R^2}\right)  =  \frac{D}{16\pi \sigma R^2 \left(R-\frac{3M}{2}\right)}. \label{sffe}
\end{eqnarray}
From equations (\ref{fffe2}) and (\ref{sffe}) we can  find that $M$ is given by

\begin{equation}
    M=\frac{6 R^3}{18 R^2+1}.
\end{equation}

\begin{figure}
    \centering
    \includegraphics[scale=0.25]{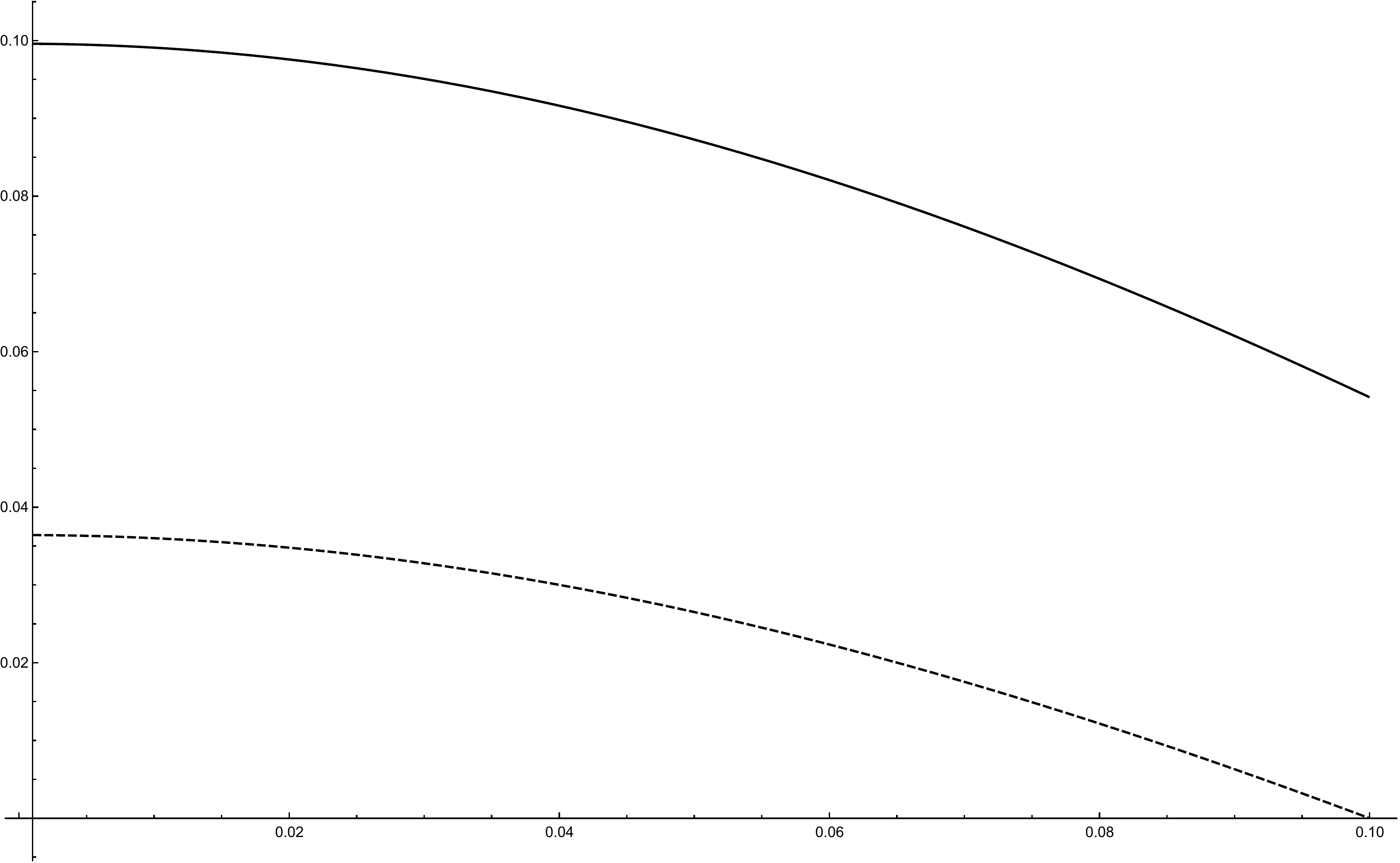}
    \caption{Qualitative comparison of the pressure for a distribution of $R=0.1$ in the brane world model (continuous curve) with the general relativistic case (dashed curve) when $p(R)=0$.}
    \label{fig:pressurescomp1}
\end{figure}

\begin{figure}
    \centering
    \includegraphics[scale=0.25]{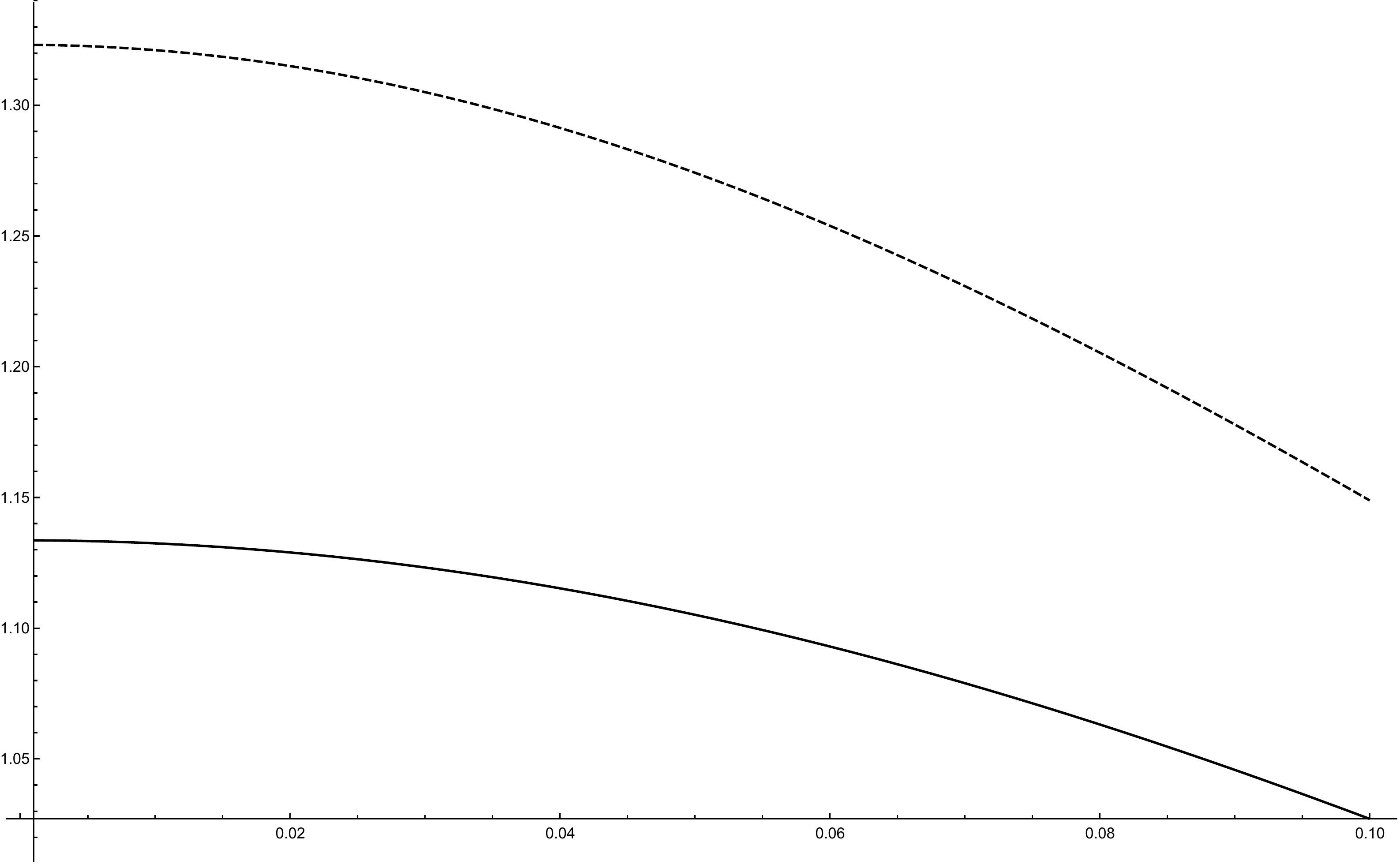}
    \caption{Qualitative comparison of the energy density for a distribution of $R=0.1$ in the brane world model (continuous curve) with the general relativistic case (dashed curve) when $p(R)=0$.}
    \label{fig:density1}
\end{figure}

\begin{figure}
    \centering
    \includegraphics[scale=0.25]{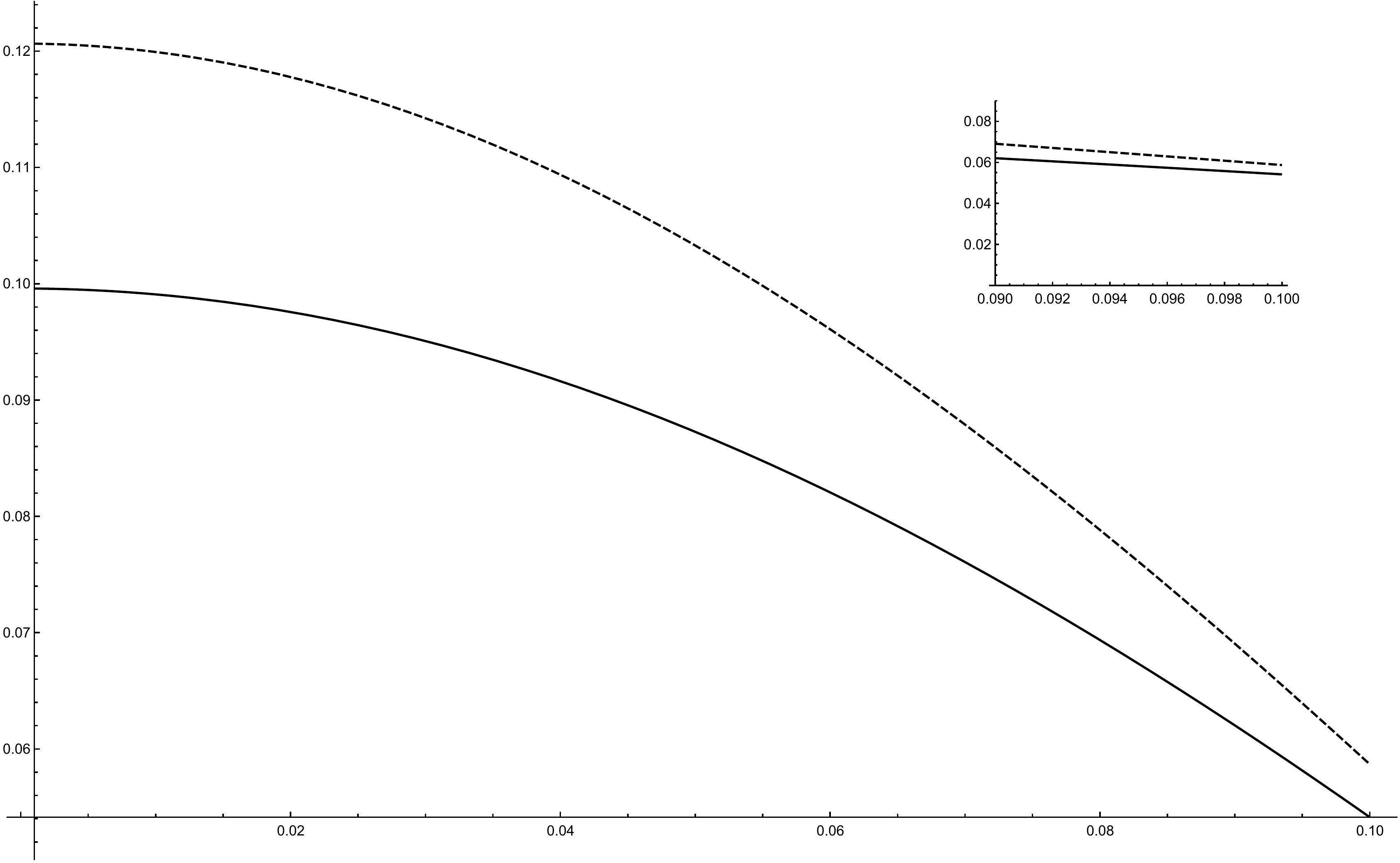}
    \caption{Qualitative behavior of the radial pressure (continuous curve) and tangential pressure (dashed curve) $R=0.1$ when $p(R)=0$.}
    \label{fig:PrPt1}
\end{figure}

\begin{figure}
    \centering
    \includegraphics[scale=0.25]{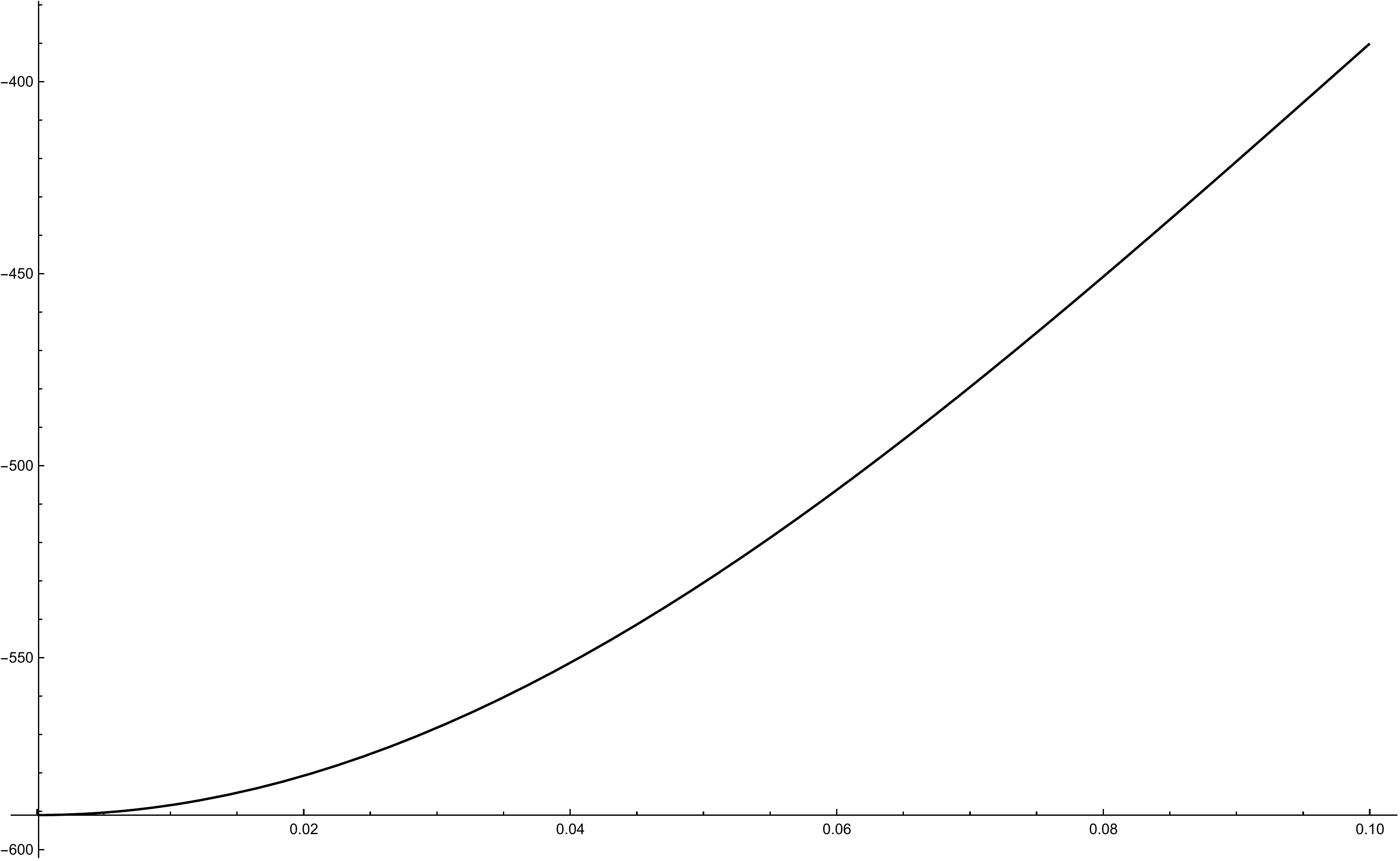}
    \caption{The scalar function $\mathcal{U}$ for $R=0.1$ when $p(R)=0$.}
    \label{fig:U1}
\end{figure}

\begin{figure}
    \centering
    \includegraphics[scale=0.25]{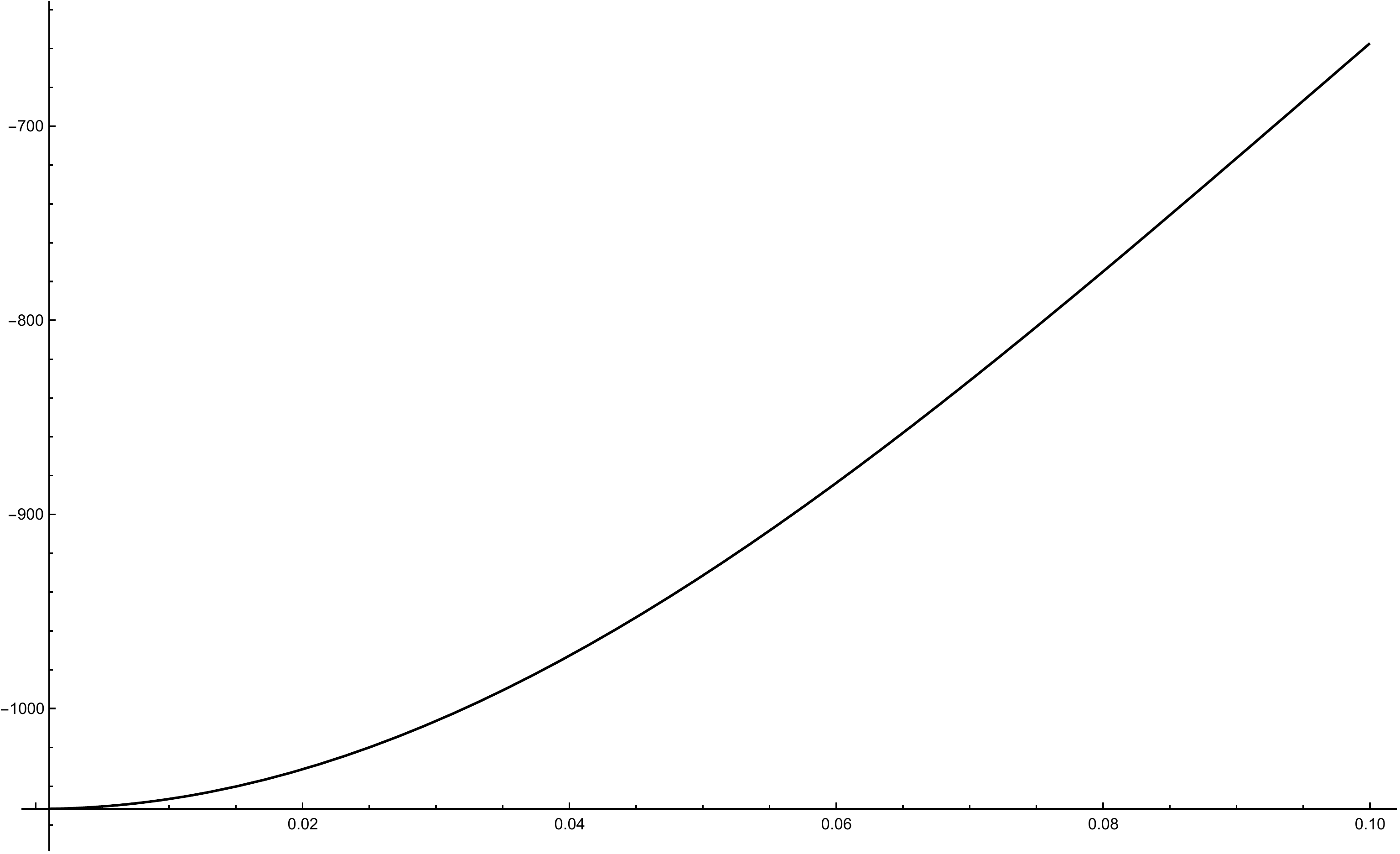}
    \caption{The Weyl function $\mathcal{P}$ for $R=0.1$ when $p(R)=0$.}
    \label{fig:P1}
\end{figure}

Thus, for a given value of $R$ and $\sigma$ we can obtain the value of $M$. Then it is possible to compute $D$ and $B$ from equations $(\ref{fffe2})$ and $\ref{fffe}$, which lead to

\begin{eqnarray}
    D & = & \frac{(3 M-2 R)}{(2 M-1) \left(18 R^2+1\right)}\Big(36 a R^2 M+2 a M-12 a R^3  \nonumber \\ & +&  18 R^3 f^*(R)+R f^*(R)\Big), \\
    B^2 & = & \frac{R-2M}{R(1+6R^2)}.
\end{eqnarray}
In order to give an example of the qualitative behavior of the distribution (that satisfied all the physical acceptability conditions) we choose $R=0.1$ and $\sigma=5$. The results are shown in the figures (\ref{fig:pressurescomp1})-(\ref{fig:P1}).

\textbf{The case with $p(R)\not= 0$.} \\

Now, if we drop the condition (\ref{sffgr}) it is possible to match our solution with the exterior Schwarzschild vacuum. In this case the matching conditions yield

\begin{eqnarray}
&& B^2 \left(6 R^2+1\right)  =  \left(1-\frac{2M}{R}\right), \label{fffe12} \\
&& \frac{\left(6 R^2+1\right) \left(1-\frac{R^2}{C^2}\right)}{12 R^2+1} + \frac{1}{\sigma}f^{*}(R)  =  \left(1-\frac{2M}{R}\right), \label{fffe22} \\
&& \frac{6 C^2-18 R^2-1}{8 F^2 \left(12 \pi  R^2+\pi \right)}+\frac{f^*(R)}{8\pi \sigma}\left(\frac{12}{6 R^2+1}+\frac{1}{R^2}\right)  =  0. \label{sffe2}
\end{eqnarray}
Then, using Eqs. (\ref{sffe2}) we can obtain

\begin{equation}
    K_1(R) F^4 + K_2(R) F^2+ K_3 (R)=0 \label{efF}
\end{equation}
where

\begin{eqnarray}
K_1(R) & = & 216 \left(18 R^2+1\right) \left(12 R^2+1\right)^3 \nonumber \\ &\times & \log \left(\sqrt{9 R^2+1}+3 R\right) \nonumber \\ &+& 144 \sqrt{3} \left(18 R^2+1\right) \left(12 R^2+1\right)^3 \nonumber \\ & \times & \tan ^{-1}\left(\frac{R}{\sqrt{3 R^2+\frac{1}{3}}}\right) \nonumber \\ &+& 216 R \Big( 41472 \pi \sigma  R^8+144 (80 \pi \sigma -27) R^6 \nonumber \\ &+& 12 (88 \pi \sigma -123) R^4 \nonumber \\ &+& 32 (\pi \sigma -5) R^2-5\Big) \sqrt{9 R^2+1}, \\
K_2(R) & = & 24 \sqrt{3} \left(12 R^2+1\right)^3 \left(18 R^2+1\right)\nonumber \\ &\times & \tan ^{-1}\left(\frac{R}{\sqrt{3 R^2+\frac{1}{3}}}\right) \nonumber \\ &-& 72 R \sqrt{9 R^2+1} \left(18 R^2+1\right) \Big(20736 \pi \sigma  R^8 \nonumber \\&+& 144 (40 \pi \sigma  -9) R^6 +12 (44 \pi \sigma -27) R^4 \nonumber \\ &+& 2 (8 \pi \sigma -5) R^2+1\Big), \\
K_3(R) & = & -3 \sqrt{9 R^2+1} R+\sqrt{3} \left(12 R^2+1\right)^3 \nonumber \\ &\times & \left(18 R^2+1\right) \tan ^{-1}\left(\frac{R}{\sqrt{3 R^2+\frac{1}{3}}}\right) \nonumber \\ &-& 2519424 \sqrt{9 R^2+1} R^{11}-769824 \sqrt{9 R^2+1} R^9 \nonumber \\ &-& 81648 \sqrt{9 R^2+1} R^7 -3996 \sqrt{9 R^2+1} R^5 \nonumber \\ &-& 132 \sqrt{9 R^2+1} R^3.
\end{eqnarray}
Thus, for a given value of $R$, it is possible to find a value for $F$ and then, using Eqs. (\ref{fffe12})-(\ref{fffe22}), we can find an expression for $M$ and $B$

\begin{eqnarray}
M & = & \frac{R^3 \left(6 F^2+6 R^2+1\right)}{2 F^2 \left(12 R^2+1\right)}-\frac{R}{2\sigma}f^*(R),  \label{mc2}\\
B^2 & = & \frac{R-2M}{R(1+6R^2)}.
\end{eqnarray}
Now in order to give an example of the qualitative behavior (which satisfies all the physical acceptability conditions) in this case we choose the values $R=0.1$ and $\sigma=5$. The results are presented in the figures.

\begin{figure}
    \centering
    \includegraphics[scale=0.25]{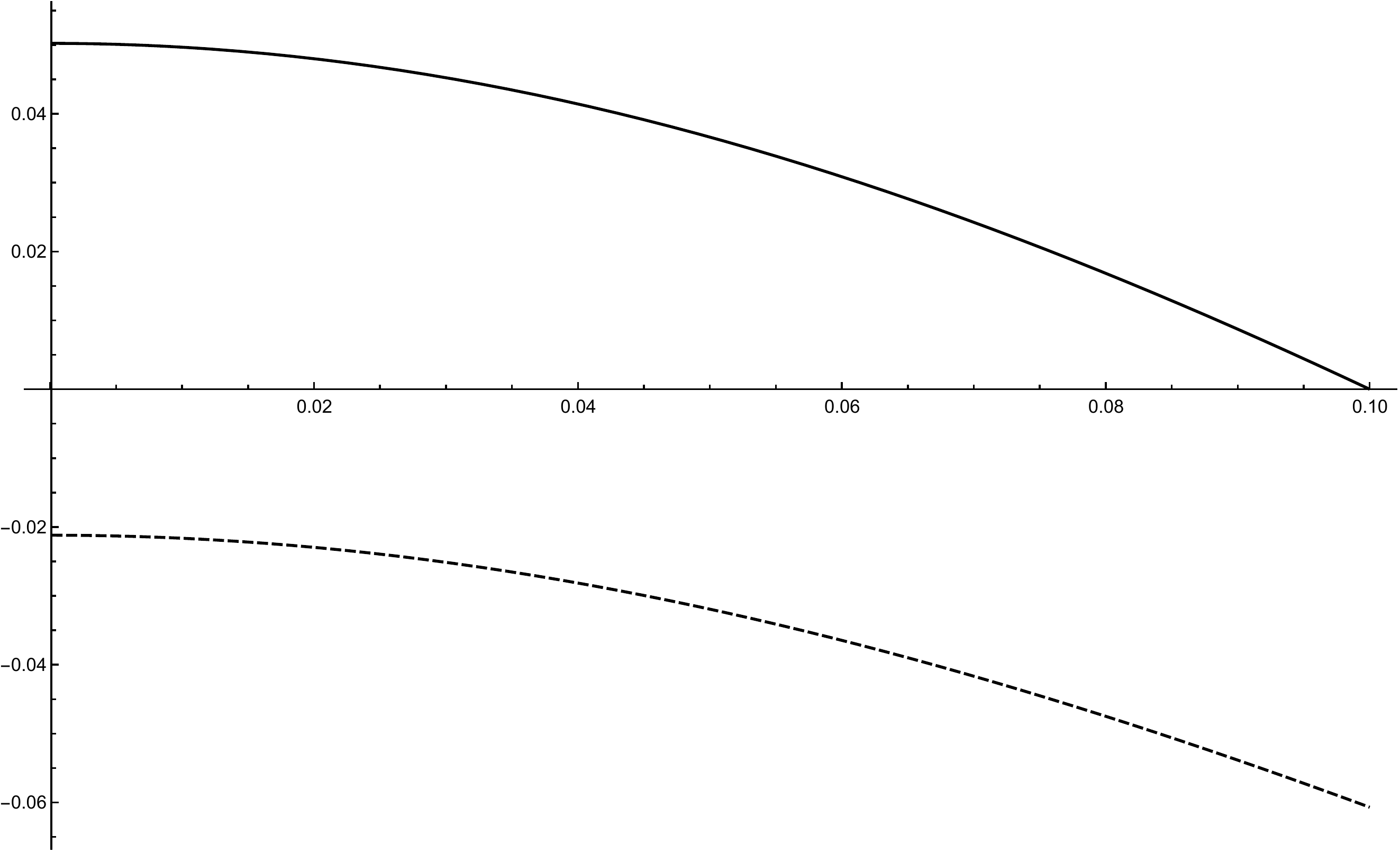}
    \caption{Qualitative comparison of the pressure for a distribution of $R=0.1$ in the brane world model (continuous curve) with the general relativistic case (dashed curve) when $p(R)\not=0$.}
    \label{fig:pressurescomp2}
\end{figure}

\begin{figure}
    \centering
    \includegraphics[scale=0.25]{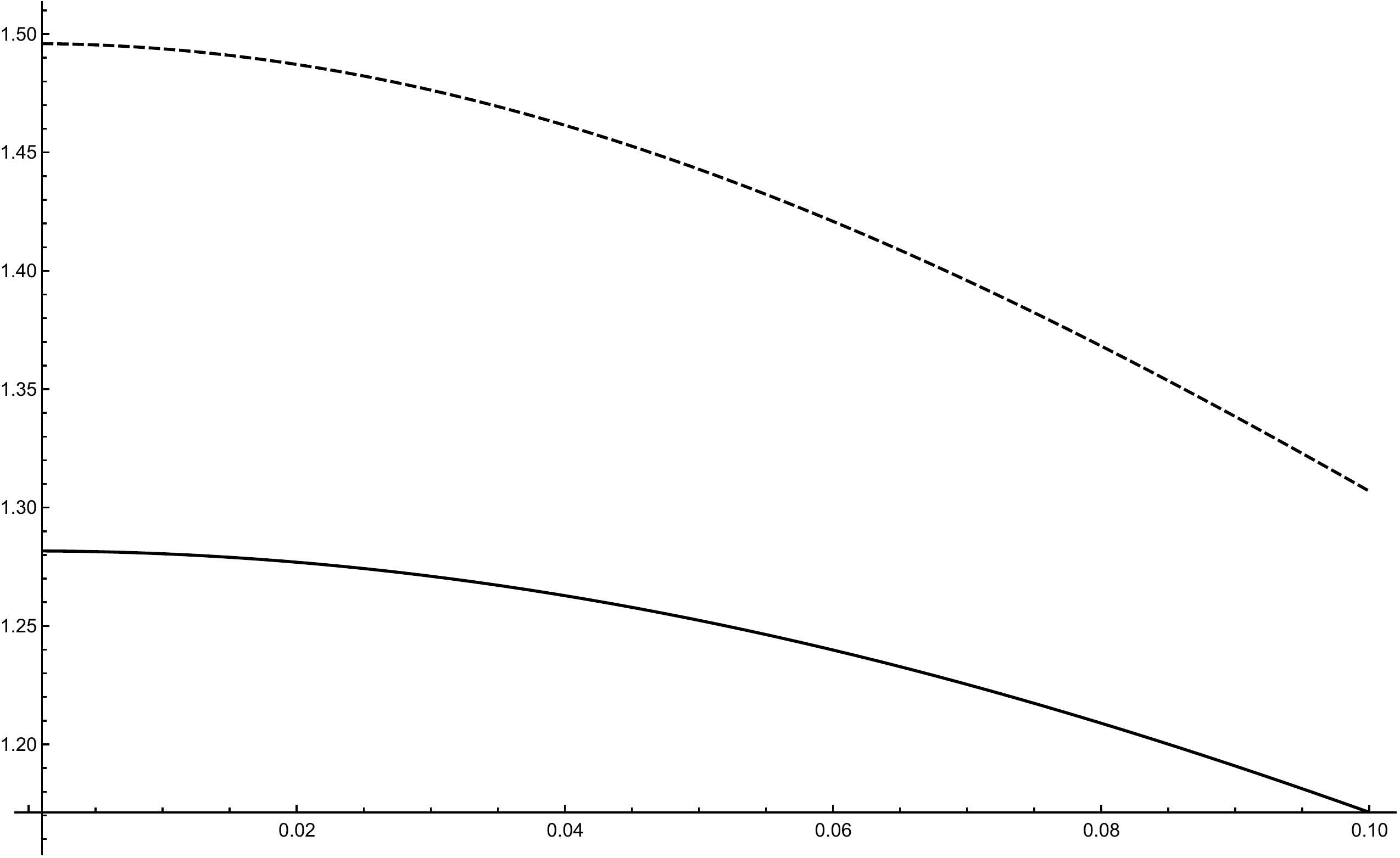}
    \caption{Qualitative comparison of the energy density for a distribution of $R=0.1$ in the brane world model (continuous curve) with the general relativistic case (dashed curve) when $p(R)\not=0$.}
    \label{fig:density2}
\end{figure}

\begin{figure}
    \centering
    \includegraphics[scale=0.25]{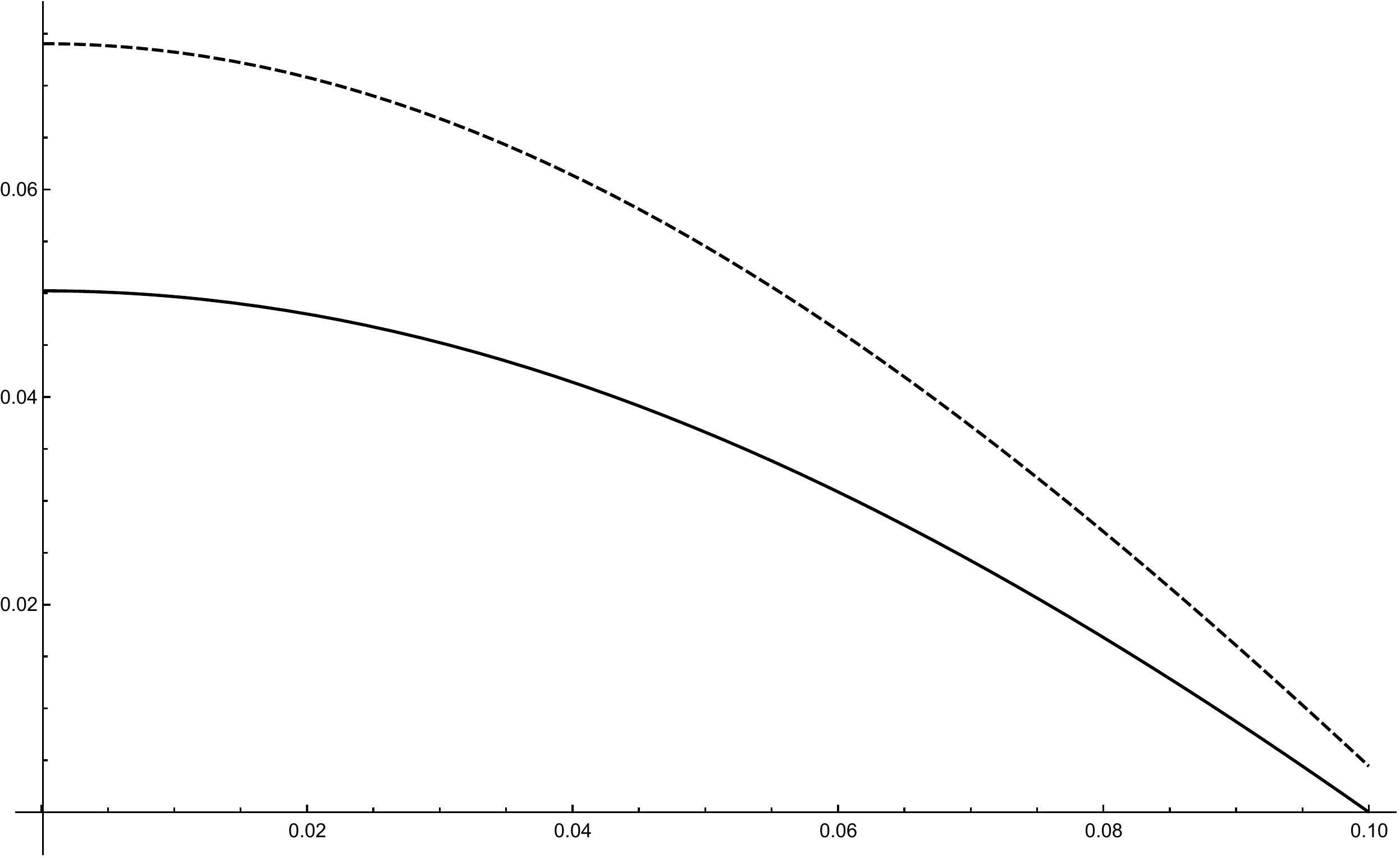}
    \caption{Qualitative behavior of the radial pressure (continuous curve) and tangential pressure (dashed curve) $R=0.1$ when $p(R)\not=0$.}
    \label{fig:PrPt2}
\end{figure}

\begin{figure}
    \centering
    \includegraphics[scale=0.25]{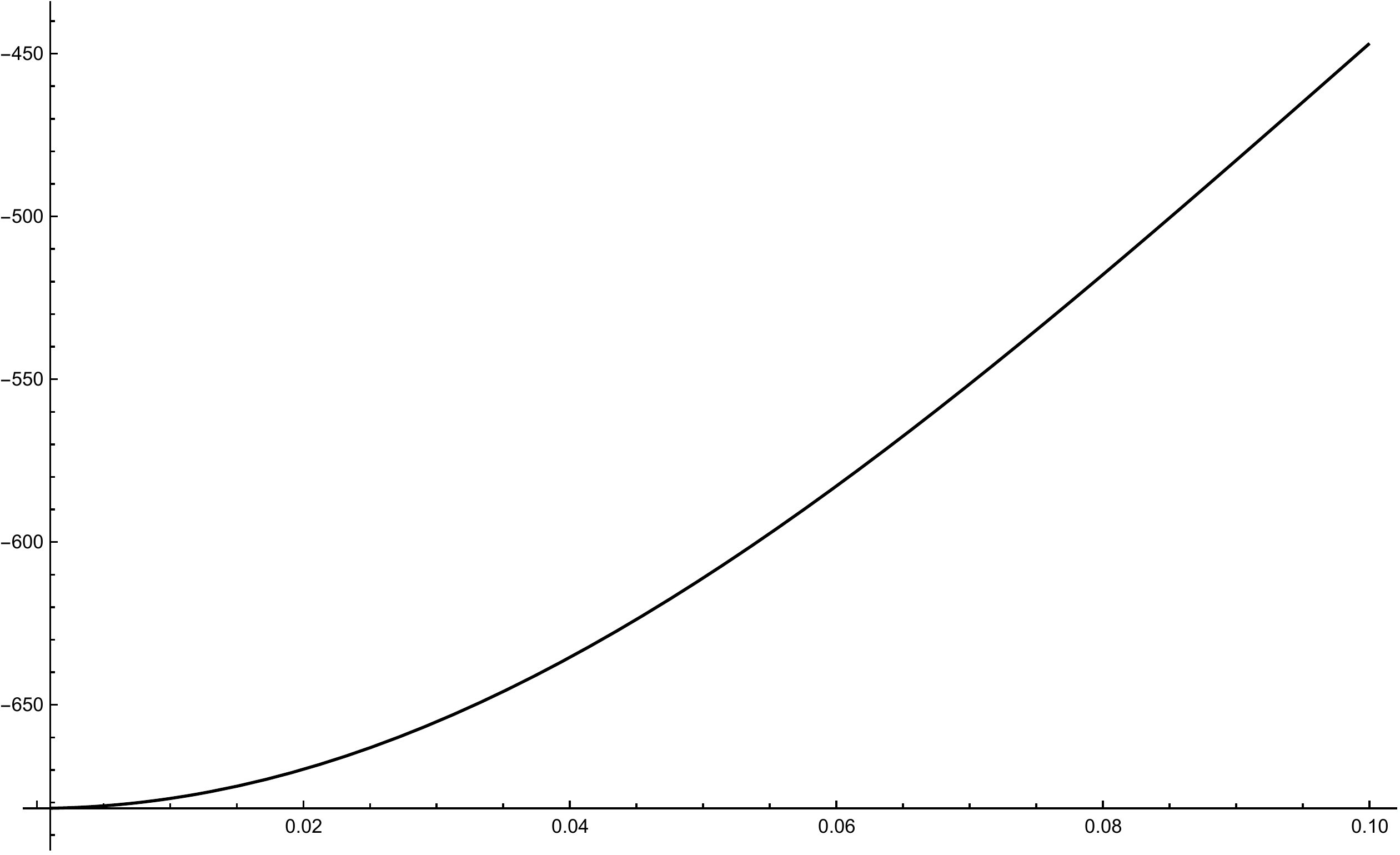}
    \caption{The scalar function $\mathcal{U}$ for $R=0.1$ when $p(R)\not=0$.}
    \label{fig:U2}
\end{figure}

\begin{figure}
    \centering
    \includegraphics[scale=0.25]{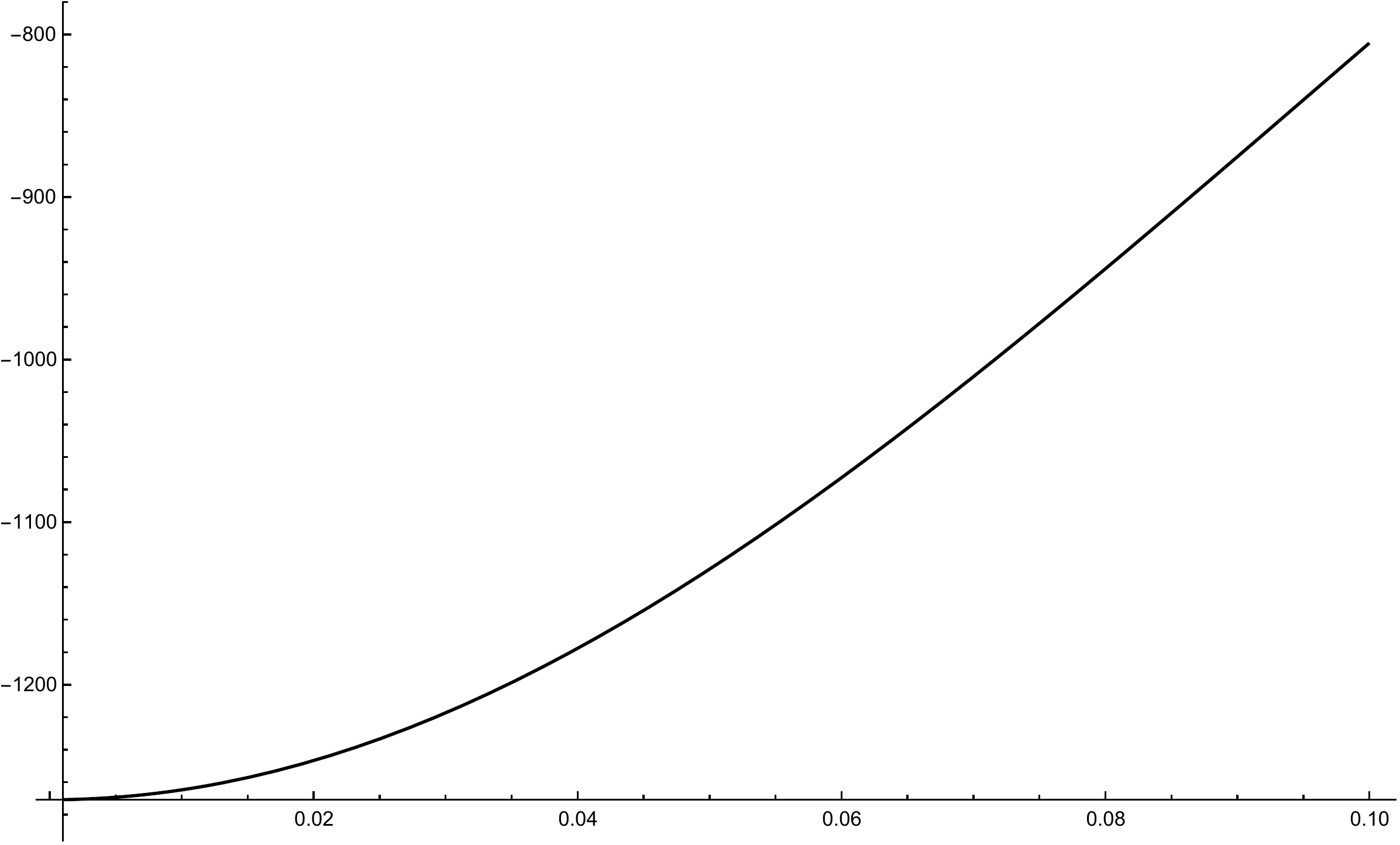}
    \caption{The Weyl function $\mathcal{P}$ for $R=0.1$ when $p(R)\not=0$.}
    \label{fig:P2}
\end{figure}

\subsubsection{The anisotropic case}

As we mention before the MGD method is a very powerful method to extend the GR solutions to the RSBW scenario, however the previous version of this method is restricted to isotropic matter distributions. We extend this results to include the anisotropic case. In order to show how this work let us consider a compact object sustained only with tangential pressure given by

\begin{eqnarray}
e^\nu &=& B^2\left(1+\frac{r^2}{A^2}\right) \label{pr1}, \\
e^{-\lambda}&=&\frac{A^2+r^2}{A^2+3r^2}, \label{pr2} \\
\rho &=& \frac{6(A^2+r^2)}{k^2(A^2+3r^2)^2} \label{pr3} \\
p_t &=&\frac{3r^2}{k^2(A^2+3r^2)^2}, \label{pr4} \\
p_r &=& 0 \label{pr5}.
\end{eqnarray}   
This solution can be interpreted (among other interpretations) as a cluster of particles moving in randomly oriented circular orbits. Now the Eqs. (\ref{pr1})-(\ref{pr5}) represent a solution for the system (\ref{2.142})-(\ref{2.162}), and then, we can use the MGD method to find its RSBW version. Thus, introducing theses expressions in (\ref{tawh2}) we obtain

\begin{eqnarray}
f^*(r) &=& \frac{\left(A^2+r^2\right)} {16 \pi  r \left(A^2+3 r^2\right)^3 \left(2 A^2+3 r^2\right)^{3/2}}\left(2 \sqrt{3} \left(A^2+3 r^2\right)^3  \right. \nonumber \\ & \times & \log \left(\sqrt{6 A^2+9 r^2}+3 r\right) \nonumber \\ &+& \sqrt{3} \left(A^2+3 r^2\right)^3 \tan ^{-1}\left(\frac{r}{\sqrt{\frac{2 A^2}{3}+r^2}}\right) \nonumber \\ &-& \left.  r \left(2 A^4+7 A^2 r^2+6 r^4\right) \sqrt{2 A^2+3 r^2}\right),
\end{eqnarray}
were $J=0$ (in Eq. (\ref{tawh2})) to avoid a singularity in the center of the distribution. 
Now, as before, in order to study the behavior of this function at the center of the distribution  we can do an expansion in power series around $r=0$,

\begin{eqnarray}
f^*(r)&=&\frac{\sqrt{\frac{3}{2}} \log \left(\sqrt{6} \sqrt{A^2}\right)}{16 \pi  \sqrt{A^2} r}-\frac{5 r \left(\sqrt{\frac{3}{2}} \log \left(\sqrt{6} \sqrt{A^2}\right)\right)}{64 \left(\pi  \left(A^2\right)^{3/2}\right)} \nonumber \\ &+&\frac{3 r^2}{4 \pi  A^4}+\frac{63 \sqrt{\frac{3}{2}} r^3 \log \left(\sqrt{6} \sqrt{A^2}\right)}{512 \pi  A^4 \sqrt{A^2}}\nonumber \\&-&\frac{93 r^4}{20 \left(\pi  A^6\right)}+O\left(r^5\right),
\end{eqnarray}
from which it is easy to see that that we need to have $A^2=1/6$ in order to ensure the regularity at the origin of the distribution. In this case 

\begin{eqnarray}
f^*&=&\frac{3 \left(6 r^2+1\right)}{16 \pi  r \left(9 r^2+1\right)^{3/2} \left(18 r^2+1\right)^3} \left(\left(18 r^2+1\right)^3 \right. \nonumber \\ &\times & \log \left(\sqrt{9 r^2+1}+3 r\right) \nonumber \\ &+&  \left(18 r^2+1\right)^3 \tan ^{-1}\left(\frac{r}{\sqrt{r^2+\frac{1}{9}}}\right) \nonumber \\ &-& \left. 6 r \left(108 r^4+21 r^2+1\right) \sqrt{9 r^2+1}\right),
\end{eqnarray}
and

\begin{eqnarray}
\mathcal{U} & = & \frac{4 \pi  \left(54 r^2+15\right) f^*(r)}{3 \left(54 r^4+15 r^2+1\right)} \nonumber \\ &-& \frac{108 \left(81 r^4+27 r^2+2\right)}{\left(18 r^2+1\right)^4}, \\
\mathcal{P} & = & \frac{2 \pi  \left(144 r^4+22 r^2+1\right) f^*(r)}{54 r^6+15 r^4+r^2} \nonumber \\ &-& \frac{54 \left(270 r^4+63 r^2+4\right)}{\left(18 r^2+1\right)^4}.
\end{eqnarray}

Now, due to the fact that $p_r=0$ for this distribution we can not match this solution with the Schwarzschild's vacuum. In this case we can use the deformation of the Schwarzschild's exterior solution from which the matching conditions have the same form of equations (\ref{fffe})-(\ref{sffe}). Thus, we also has the same form for the constants $M$, $B$ and $D$, that is

\begin{equation}
    M=\frac{6 R^3}{18 R^2+1}.
\end{equation}

\begin{eqnarray}
    D & = & \frac{(3 M-2 R)}{(2 M-1) \left(18 R^2+1\right)}\Big(36 a R^2 M+2 a M-12 a R^3  \nonumber \\ & +&  18 R^3 f^*(R)+R f^*(R)\Big), \\
    B^2 & = & \frac{R-2M}{R(1+6R^2)}.
\end{eqnarray}

\begin{figure}
    \centering
    \includegraphics[scale=0.25]{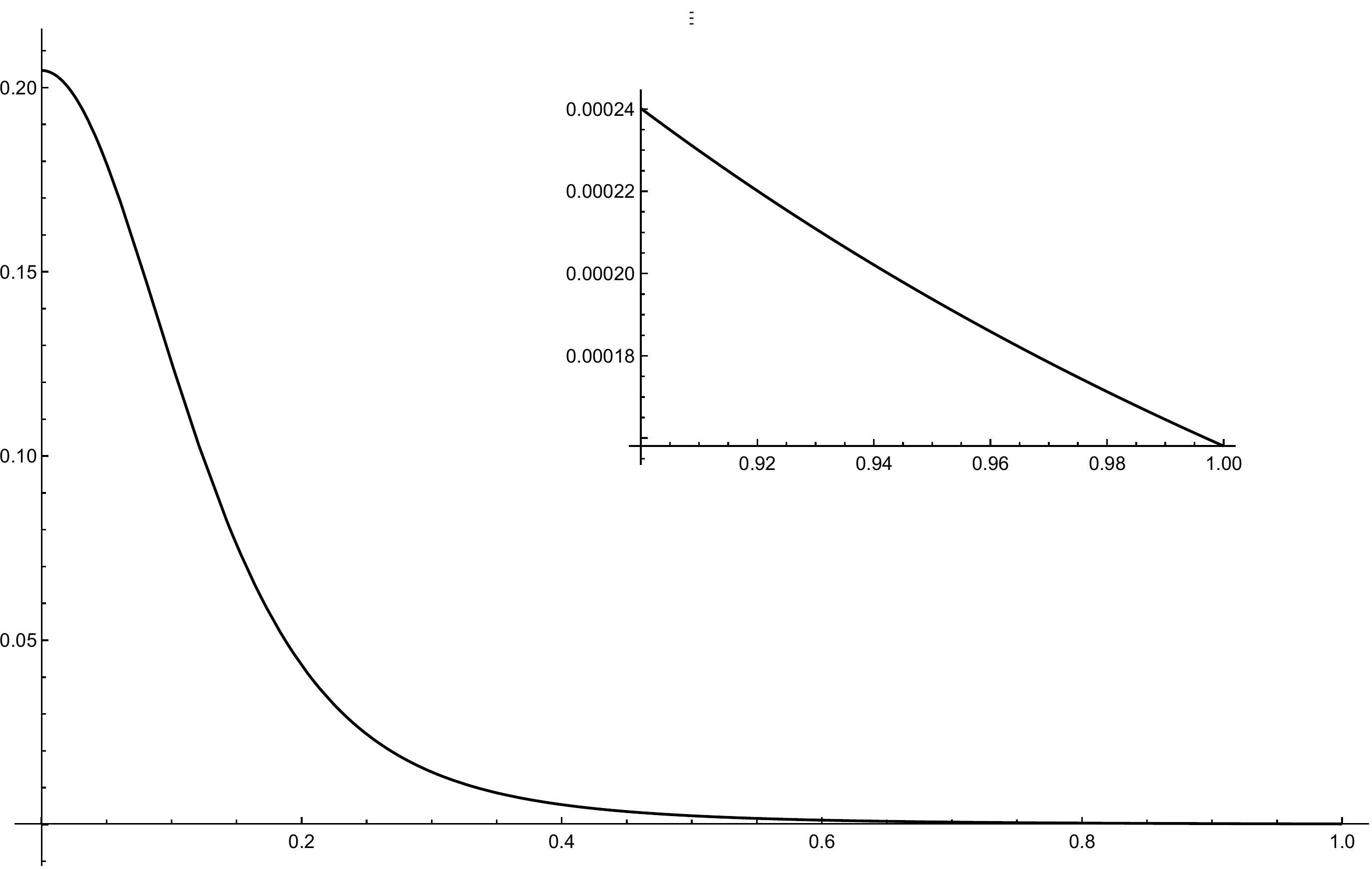}
    \caption{Qualitative behavior of the radial pressure for a distribution with $R=1$ in the brane world model.}
    \label{fig:pressurescomp23}
\end{figure}

\begin{figure}
    \centering
    \includegraphics[scale=0.25]{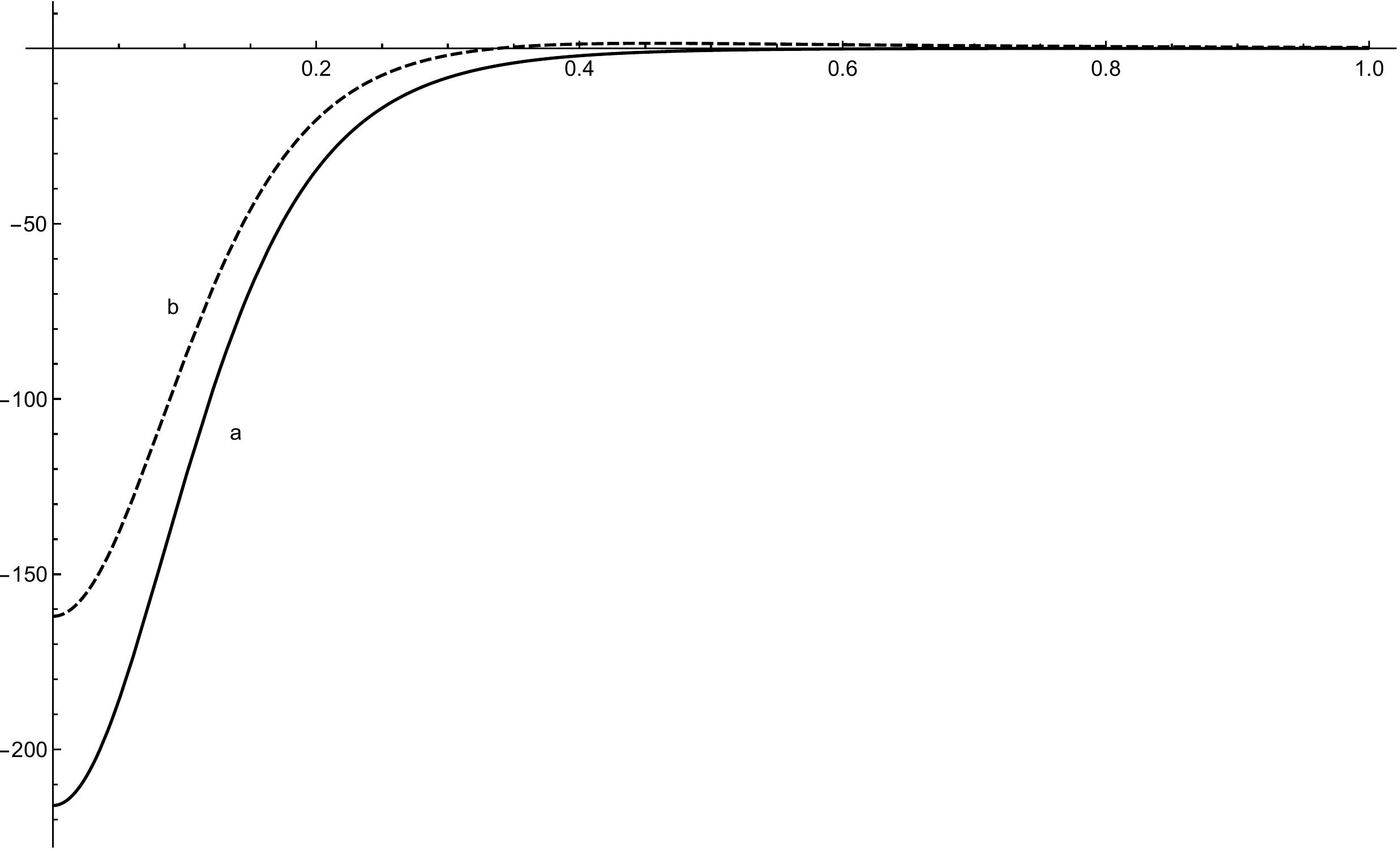}
    \caption{The scalar and Weyl functions for $R=1$.}
    \label{fig:pressurescomp133}
\end{figure}

As an example we show some results for the qualitative behavior of all relevant quantities for the values $R=1$ and $\sigma=5$ in the figures (\ref{fig:pressurescomp23})-(\ref{fig:pressurescomp133}). 

\begin{figure}
    \centering
    \includegraphics[scale=0.25]{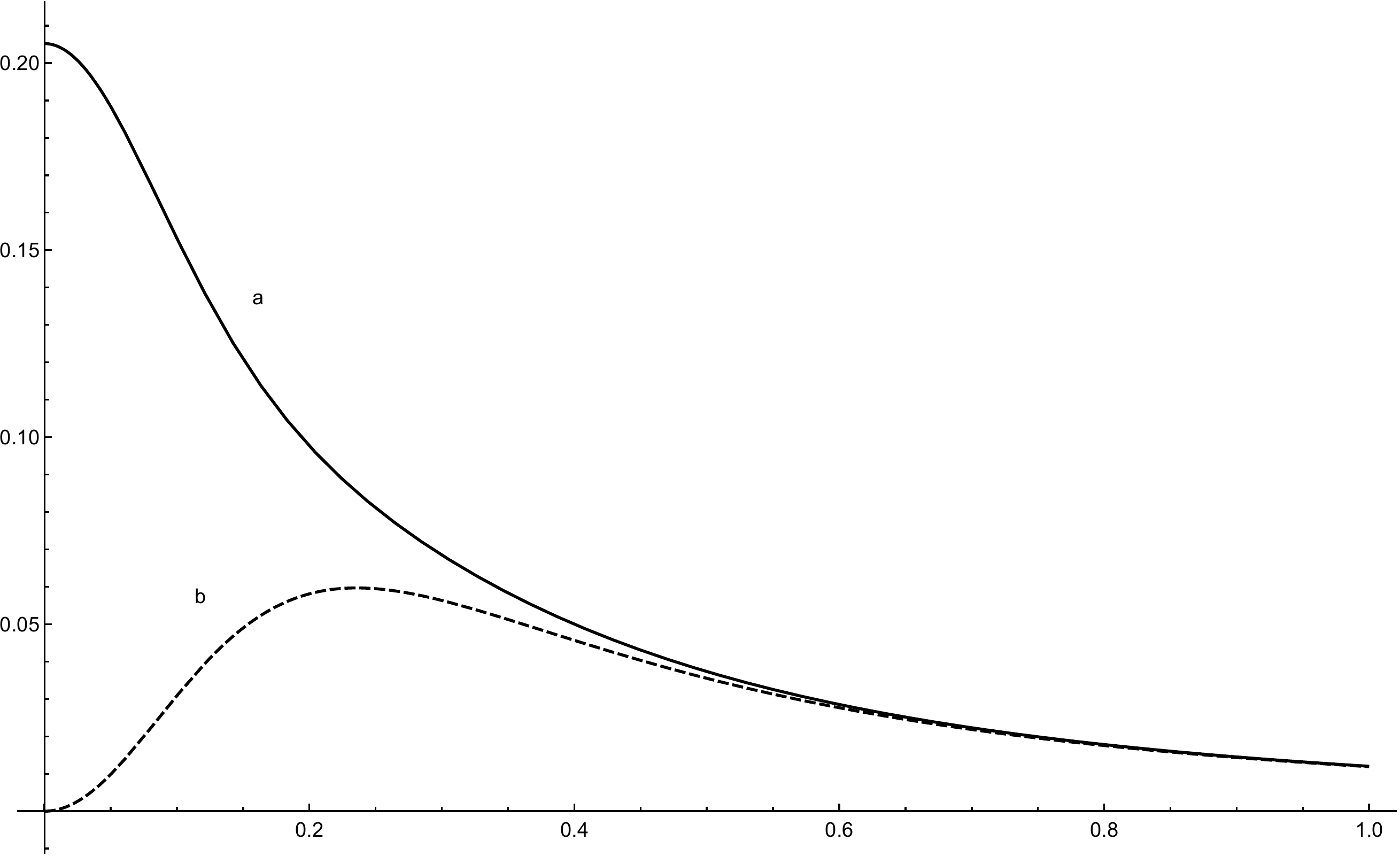}
    \caption{Qualitative comparison of the tangential pressure for a distribution with $R=1$ in the brane world model (continuous curve) with the general relativistic case (dashed curve).}
    \label{fig:pressurescomp13}
\end{figure}

\begin{figure}
    \centering
    \includegraphics[scale=0.25]{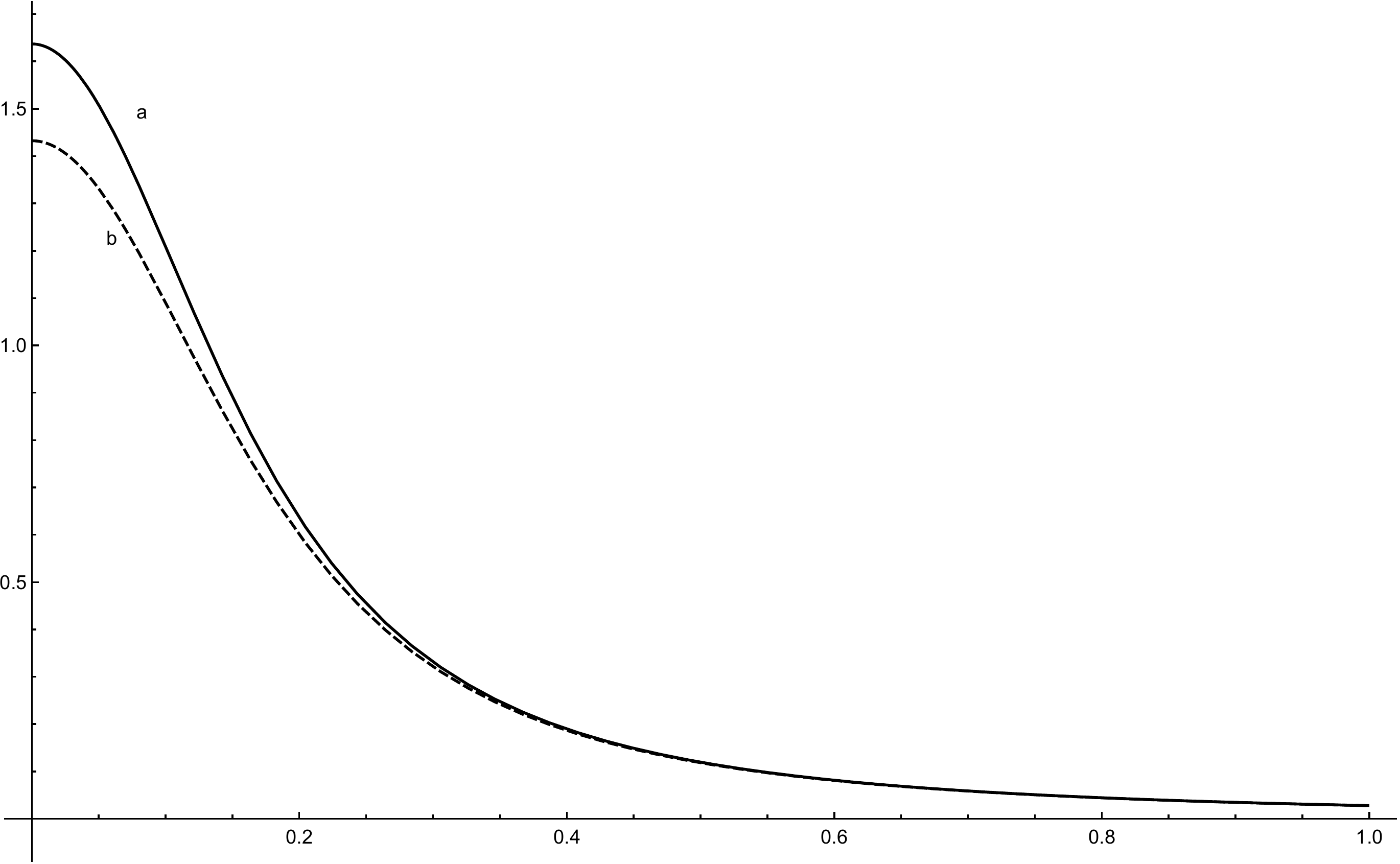}
    \caption{Qualitative comparison of the energy density for a distribution with $R=1$ in the brane world model (continuous curve) with the general relativistic case (dashed curve).}
    \label{fig:pressurescomp13}
\end{figure}

\section{Conclusions}

In this paper we presented the general formalism to obtain analytical solutions of the effective Einstein equations in the RSBW model, using the MGD-decoupling method. In particular, we used this approach to study new black holes solutions in two different ways. The first one was by deforming the exterior Schwarzschild vacuum, in this case we obtained a solution that was reported in Ref.~\cite{Germani:2001du}. Being this the only solution we can find deforming the Schwarzschild vacuum, it shows the limitations of the MGD decoupling method in this case. Then, in order to obtain more black holes solutions starting with the  Schwarzschild solution it is  necessary to use the extended version of the MGD-decoupling method. The second approach that we used to study the RSBW solutions, was starting with the known tidial charge black hole solution of the RSBW. Then, using the MGD-decoupling with different conditions over the source $\theta_{\mu \nu}$ we have been able to find three new black holes solutions. However, by requiring that the new solutions be asymptotically flat, then we concluded that the source $\theta_{\mu \nu}$ can not have isotropic pressures ($\theta^1_1=\theta^2_2=\theta^3_3$) or tangential null pressure ($\theta^2_2=0$).

We also presented how to obtain internal analytical solutions of the effective Einstein's field equations. In that sense we showed that every internal anisotropic solution of the field equations in General Relativity can be extended to the brane world by using the MGD-decoupling. In order to give examples on how the method works, we used two different GR internal solutions (one isotropic distribution, the Tolman IV internal solution and other, anisotropic in pressures), and using the steps presented in this work we were able to find new solutions in the brane world scenario.

Furthermore, we discussed the matching conditions corresponding to the obtained solutions. We found two different possibilities in the isotropic domain; in fact, we used two different approaches for the matching conditions. The first one was assuming that the physical pressure is zero at the surface of the distribution, that is $p(R)=0$. In this case we showed that our internal solution could not be coupled to the Schwarzschild's vacuum. Instead of this we used the external solution which was found at the beginning of the section 4. The second one approach was based on requiring that the effective radial pressure be zero at the surface and in this case we were able to found how to match our solution with the Schwarzschild's vacuum. For the anisotropic case we were only be able to match our solution to the deformation of Schwarzschild's vacuum. 

In order to obtain more physically acceptable solutions in the RSBW we may choose other solutions of Einstein's equations in General Relativity and then follow the procedure given here. We may also start with the same initial solutions of General Relativity presented here and then use the extended version of MGD-decoupling method, in which we can also consider a temporal deformation of the metric.

\section{Acknowledgements}

P. L. wants to say thanks for financial support received by CONICYT PFCHA/DOCTORADO BECAS CHILE/2019 -2119051, the Project ANT1856 and Cem 18-02 Project of Universidad de Antofagasta. 
A. S. was partially supported by Project Fondecyt 1161192 (Chile) and also by the MINEDUC-UA project, code ANT 1855.

\section*{Conflict of interest}

The authors declare no potential conflict of interests.

\balance

\providecommand{\url}[1]{\texttt{#1}}
\providecommand{\urlprefix}{}
\providecommand{\foreignlanguage}[2]{#2}
\providecommand{\Capitalize}[1]{\uppercase{#1}}
\providecommand{\capitalize}[1]{\expandafter\Capitalize#1}
\providecommand{\bibliographycite}[1]{\cite{#1}}
\providecommand{\bbland}{and}
\providecommand{\bblchap}{chap.}
\providecommand{\bblchapter}{chapter}
\providecommand{\bbletal}{et~al.}
\providecommand{\bbleditors}{editors}
\providecommand{\bbleds}{eds: }
\providecommand{\bbleditor}{editor}
\providecommand{\bbled}{ed.}
\providecommand{\bbledition}{edition}
\providecommand{\bbledn}{ed.}
\providecommand{\bbleidp}{page}
\providecommand{\bbleidpp}{pages}
\providecommand{\bblerratum}{erratum}
\providecommand{\bblin}{in}
\providecommand{\bblmthesis}{Master's thesis}
\providecommand{\bblno}{no.}
\providecommand{\bblnumber}{number}
\providecommand{\bblof}{of}
\providecommand{\bblpage}{page}
\providecommand{\bblpages}{pages}
\providecommand{\bblp}{p}
\providecommand{\bblphdthesis}{Ph.D. thesis}
\providecommand{\bblpp}{pp}
\providecommand{\bbltechrep}{}
\providecommand{\bbltechreport}{Technical Report}
\providecommand{\bblvolume}{volume}
\providecommand{\bblvol}{Vol.}
\providecommand{\bbljan}{January}
\providecommand{\bblfeb}{February}
\providecommand{\bblmar}{March}
\providecommand{\bblapr}{April}
\providecommand{\bblmay}{May}
\providecommand{\bbljun}{June}
\providecommand{\bbljul}{July}
\providecommand{\bblaug}{August}
\providecommand{\bblsep}{September}
\providecommand{\bbloct}{October}
\providecommand{\bblnov}{November}
\providecommand{\bbldec}{December}
\providecommand{\bblfirst}{First}
\providecommand{\bblfirsto}{1st}
\providecommand{\bblsecond}{Second}
\providecommand{\bblsecondo}{2nd}
\providecommand{\bblthird}{Third}
\providecommand{\bblthirdo}{3rd}
\providecommand{\bblfourth}{Fourth}
\providecommand{\bblfourtho}{4th}
\providecommand{\bblfifth}{Fifth}
\providecommand{\bblfiftho}{5th}
\providecommand{\bblst}{st}
\providecommand{\bblnd}{nd}
\providecommand{\bblrd}{rd}
\providecommand{\bblth}{th}


\end{document}